\documentclass[10pt,aps,prc,twocolumn,amssymb,showpacs,floatfix]{revtex4-1}
\pdfoutput=1
\usepackage{amsmath}
\usepackage{amssymb}
\usepackage[british]{babel}
\usepackage{graphicx}
\usepackage{enumerate}
\usepackage{epstopdf}
\usepackage[breaklinks,colorlinks]{hyperref}

\hypersetup{
  citecolor=blue
}
\usepackage{graphicx}

\newcommand{\half}[0]{\frac{1}{2}}
\newcommand{\thrhalf}[0]{\frac{3}{2}}
\newcommand{\deriv}[2]{\frac{d #1}{d #2}}

\newcommand{\pderiv}[2]{\frac{\partial #1}{\partial #2}}
\newcommand{\pderivtwo}[2]{\frac{\partial^2 #1}{\partial #2^2}}
\newcommand{\bref}[1]{(\ref{#1})}
\newcommand{\brac}[1]{\left( #1\right)}
\newcommand{\sbrac}[1]{\left[ #1\right]}

\newcommand{\hl}[0]{h_l^{(1)}}

\newcommand{\nucl}[3]{
\ensuremath{
\phantom{\ensuremath{^{#1}_{#2}}}
\llap{\ensuremath{^{#1}}}
\llap{\ensuremath{_{\rule{0pt}{.75em}#2}}}
\mbox{#3}
}
}

\begin{document}

\title{Continuum time-dependent Hartree-Fock for giant resonances in spherical nuclei}
\author{C. I. Pardi}
\affiliation{Department of Physics, University of Surrey, Guildford, Surrey, GU2 7XH, United Kingdom}
\author{P. D. Stevenson}
\affiliation{Department of Physics, University of Surrey, Guildford, Surrey, GU2 7XH, United Kingdom}
\date{\today}

\begin{abstract}
This paper deals with the solution of the spherically symmetric time-dependent Hartree-Fock approximation applied in the case of nuclear giant monopole resonances in the small and large amplitude regimes. The problem is spatially unbounded as the resonance state is in the continuum.  The practical requirement to perform the calculation in a finite-sized spatial region results in a difficulty with the spatial boundary conditions. Here we propose a absorbing boundary condition scheme to handle the conflict. The derivation, via a Laplace transform method, and implementation is described. The accuracy and efficiency of the scheme is tested and the results presented to support the case that they are a effective way of handling the artificial boundary.
\end{abstract}

\maketitle
 
\section{\label{sec:Intro}Introduction}

It occurs in many areas of physics that the time-evolution of a spatially unbounded system is required to be analysed. Such systems have been studied in many fields of physics involving wave propagation, spanning areas such as laser physics and gravitational waves, \cite{springerlink:10.1134/S1054660X10050063,symABC,Lau2004376,alpert2000}. Examples occur in nuclear physics and we analyse such a case in the present work.

The particular physical phenomenon being studied here is the nuclear giant monopole resonance. It is well known that these are above the particle decay threshold \cite{giantResFHMNE}, so that one allowed decay mode involves the expulsion of one or more nucleons from the nucleus. A time-dependent simulation of such a decay will involve the spatial region in which the nuclear wavefunction is non-negligible becoming larger and larger as time goes on. 

One way of analysing this sort of system is via the time-dependent Hartree-Fock (TDHF) approximation, that reduces the many-body interaction to a simpler mean field one. The simplification however still does not allow analytic solutions to be gained but allows for numerical analysis to be applied and the computational cost to be manageable. 

A common numerical implementation is to discretise the equations using time and space grids employing finite difference methods.  Here, a non-trivial problem occurs because the boundary of the finite grids impose an artificial boundary into the solution. As the outgoing wave condition for  the Hartree-Fock equation is evaluated at infinity it cannot straightforwardly be applied directly. Enforcing the wrong boundary conditions results in the solution becoming incorrect for the time after the emitted particles have reached the artificial boundary and so it can be important that the boundary is handled properly.

There are various methods available that aim to simulate or circumvent the application of the outgoing wave condition\cite{giantRes,springerlink:10.1140/epjad/i2005-06-052-x}. The most crude is just to apply a reflecting boundary sufficiently far away so that the matter being emitted does not reach it within the time of the calculation. This works and reflecting boundaries can be easily implemented but the major drawback is that one needs an increasing number of grid points in space as one wants to evolve further in time.  Eventually, this becomes computationally unfeasible.

Other methods include absorbing potentials and masking functions.  These allow the artificial boundary to be placed closer to the nucleus but generally have to be tuned to each particular case and do not in general approximate the outgoing wave condition perfectly.

Here, we present a method of implementing exact boundary conditions \cite{springerlink:10.1134/S1054660X10050063,symABC}. These rely on choosing the artificial boundary such that the potential outside of it has a simple form, so that the propogation of waves in the exterior region does not have to be dealt with explicitly.

In solving the TDHF equations, a simplified Skyrme interaction is used in the implementation which reproduces the magic numbers needed for $\nucl{4}{2}{He}$, $\nucl{16}{8}{O}$ and $\nucl{40}{20}{Ca}$ to be seen without the complexity of the full interaction \cite{PhysRevC.60.044302}, as a reasonable proof-of-concept. Spherical symmetry is also assumed inside and outside of the artificial boundary. The calculations involves one, in the case of $\nucl{4}{2}{He}$, or more, in the cases of $\nucl{16}{8}{O}$ and $\nucl{40}{20}{Ca}$, different forms of differential equation, each of which requires its own absorbing boundary condition to be applied. Here some continuous absorbing boundary conditions are used. Other types of absorbing boundary are fully-discrete \cite{Antoine2003157} and semi-discrete\cite{1dABC} but are not described here. A review of the various absorbing boundary conditions can be found in \cite{abcreview}.

The structure of this paper is as follows: Section \ref{sec:gmr} gives a brief summary of nuclear giant monopole resonances; sections \ref{sec:ProbInt} and \ref{sec:intDisc} describe the Hartree-Fock approximation, the first the theory and the second its discretization and implementation; section \ref{sec:probExt} and \ref{sec:boundDisc} describe the exterior problem and the absorbing boundary conditions; sections \ref{sec:ResultsAndTesting1}, \ref{sec:ResultsAndTesting2} and \ref{sec:ResultsAndTesting3} show the testing and results of our implementation which includes a short analysis of the errors caused by the discretization and strength functions for $\nucl{4}{2}{He}$, $\nucl{16}{8}{O}$ and $\nucl{40}{20}{Ca}$, and results with large-amplitude excitation.  We end with some concluding remarks.

\section{\label{sec:gmr}Giant Monopole Resonances}

Giant resonances are collective modes of excitation of finite fermionic systems \cite{bertschbroglia}.  The first evidence for their existence in atomic nuclei came in 1937, with a theoretical description and systematic experimental study coming in the next decade \cite{harakeh}.  While the first studies excited the electric isovector dipole resonance, in which protons and neutrons oscillated out of phase with each other due to the dominance of the E1 component of the photon field, other giant resonances were discovered later.  In particular, the isoscalar giant monopole resonance (GMR) was definitively reported in 1977 \cite{Harakeh1977}.

The GMR, as a compression mode, probes the nuclear equation of state \cite{Blaizot1980}, and is therefore useful in constraining nuclear models \cite{Dutra2012}.  As a spherically-symmetric excitation, it is the first port of call for testing new theoretical methods, as the symmetry renders many types of calculation more simple.  In particular, methods based on Time-Dependent Hartree-Fock have turned to giant monopole resonances in spherical doubly-magic nuclei as a proving ground \cite{Vautherin1979,Stringari1979,Lacroix1998,Wu1999,Almehed2005,Almehed2005a,Stevenson2010}. 

The present paper is written in that spirit, employing the simplified $t_0$-$t_3$ version of the Skyrme force used in previous applications \cite{Stringari1979,Wu1999}.  While the focus of this work is on the development of the boundary conditions, and the simplified Skyrme force we use should not be expected to give good agreement with experiment, it is noted that of the three nuclei considered here, the GMR has been unambiguosly observed only in $^{40}$Ca \cite{Brandenburg1983}, though the nature of giant resonances in general in nuclei as light as $^4$He is a subject of ongoing interest \cite{Tornow2012}.

The key observable calculated for the giant resonance is the linear response function, describing the response of the nucleus to an external perturbation \cite{NozPines}.  From this, one derives the strength function, related in turn to the experimental cross section for the reaction.  The strength function can be obtained, within TDHF, via the Fourier Transform of the time-dependent moment of the resonance mode desired \cite{Chinn1996} and we present calculations of such strength functions.  We note that the strength functions are particular sensitive to the success of implementation of the absorbing boundary conditions \cite{giantRes}, and provide a good measure of success, as well as being the physically relevant quantity.


\section{\label{sec:ProbInt}Time Dependent Hartree-Fock}

The time-dependent Hartree-Fock method originates with Dirac \cite{Dirac1930}, and became computationally viable for nuclear processes in the 1970s \cite{Bonche1976,Cusson1976,Devi1979}.  Since then it has been extensively used for calculating heavy-ion reactions \cite{Maruhn2006} and giant resonances \cite{Stevenson2004}, with increasingly sophisticated implementations of the effective interaction \cite{Umar2006,PhysRevC.86.044303}.  A full derivation of the Time-dependent Hartree-Fock equations in the case of Skyrme forces can be found in the original paper by Engel et al. \cite{Engel1975}.  In the present case, with the simplified Skyrme force, and omitting Coulomb, we note that the Time-Dependent Hartree-Fock equations can be written as a series of coupled non-linear Schr\"odinger equations of the form
\begin{equation}
i\hbar\frac{\partial\psi_\lambda(\vec{r},t)}{\partial t}=\hat{h}\psi_\lambda(\vec{r},t),\qquad\lambda=1,\ldots,A,
\end{equation}
where the Hartree-Fock Hamiltonian is given by
\begin{equation}
\hat{h} = -\frac{\hbar^2}{2m}\nabla^2 + a\rho(\vec{r},t) + b\rho^2(\vec{r},t),
\end{equation}
with $\rho(\vec{r},t)=\sum_{\lambda=1}^A\psi_\lambda^*(\vec{r},t)\psi_\lambda(\vec{r},t)$ denoting the particle density. The values of $a$ and $b$ used thoughtout this paper are taken from \cite{PhysRevC.56.857} where they have the values $-817.5$ MeV fm$^3$ and $3241.5$ MeV fm$^6$. In practice, the time-dependent Hartree-Fock equations are solved by evolving in time according to
\begin{equation}
\psi_\lambda(\vec{r},t+\Delta t) = e^{-i\Delta t \hat{h}/\hbar}\psi_\lambda(\vec{r},t)
\end{equation}
Specialisation to spherical symmetry, and details of discritisation methods, are given in the following sections, in which the details of the algorithm dealing with the boundary conditions are also given.

\section{\label{sec:intDisc}Interior Discretization}
As well as the coupled non-linear differential equations noted in the previous section, initial conditions are required, and are calculated from stationary Hartree-Fock. We first describe our method for calculating the stationary solution and then go on to the time-dependent case. In both we discetized the equations on equally spaced grids, for simplicity, though non-uniform grids can be in themselves useful in pushing the boundary far into the exterior region at an acceptable computational cost \cite{PhysRevC.71.024301}.

\subsection{Stationary Discretisation}
We start with the calculation of the initial condition, which itself is a non-linear problem. We solve it by the following iterative procedure:
\begin{eqnarray}
&\hat{H}^{(i)}_\alpha(r) Q_\alpha^{(i+1)}(r) = \lambda^{(i+1)}_\alpha Q_\alpha^{(i+1)}(r)& \\
&\hat{H}^{(i)}_\alpha(r) = -\half\pderivtwo{}{r} + \sbrac{\frac{l_\alpha(l_\alpha+1)}{2r^2} + V\left\{ \rho^{(i)}(r)\right\}}& \label{eq:hfhamil} \\
&\rho^{(0)}(r) = \frac{1}{4\pi r^2}\sum_\alpha g_\alpha |Q_\alpha^{(0)}(r)|& \label{TIHF_initialden} 
\end{eqnarray}
for $i\in\mathbb{N}_0$ and where $Q(r)=r\psi$ represents the reduced wave function, $V$ is the potential, and $l_\alpha$ the orbital angular momentum. We calculate the initial guess, $\rho^{(0)}(r,t)$, using harmonic oscillator wave-functions as the $Q^{(0)}_\alpha(r)$ in equation \bref{TIHF_initialden}.

Spatial discritisation of the equations is made on a uniformly-spaced grid, such that\begin{eqnarray}
r_m=m\Delta r \text{, } m=1,\hdots,M\text{, } \Delta r = \frac{R_{out}}{M} \label{TIHFspaceGrid}
\end{eqnarray}
where $M$ is the total number of gridpoints and $R_{out}$ is the distance from the origin to the spherical outer boundary.  The second derivative operator in (\ref{eq:hfhamil}) is treated with the three-point approximation.

We also require the wave functions at two additional points; $Q_\alpha^{(i)}(r_0)\equiv Q_\alpha^{(i)}(0)$ and $Q_\alpha^{(i)}(r_{M+1})\equiv Q_\alpha^{(i)}\brac{(M+1)\Delta r}$. Although our differential equation is not evaluated at these points, values of the wave function here are needed for the finite differencing.

Working with the reduced wave function leads to a boundary condition of $Q_\alpha^{(i)}(r_0)=0$. However, the large-$r$ boundary condition, that the wave function remain square-integrable and fall to zero strictly only at infinity and cannot be applied directly. We make use of that property that the wavefunctions for bound states decay exponentially as $r$ increases. Hence we can find a radius at which the wavefunction is zero, within a given accuracy, and so we choose  $Q_\alpha^{(i)}(r_{M+1})=0$ for the solution of the static Hartree-Fock equations.

This leaves us with a tridiagonal matrix eigenvalue problem at each iteration, which can be solved efficiently using the LAPACK subroutines.

We iterate until both the eigenvalue, $\lambda^{(i+1)}_\alpha$, and the mean square errors for each wave function,
\begin{eqnarray}
\epsilon_\alpha &=&
\bigg\lvert\langle Q^{(i+1)}_\alpha\mid\!\hat{H}^{(i)}\!\mid Q^{(i+1)}_\alpha\rangle^2 \nonumber \\
&&- \langle Q^{(i+1)}_\alpha\mid \!\brac{\hat{H}^{(i)}}^2 \!\mid Q^{(i+1)}_\alpha\rangle\bigg\rvert,
\end{eqnarray} 
have stopped changing, within machine precision, from one iteration to the next.
 
\subsection{Time-Dependent Discretisation} \label{section_IntDiscTDT}
After the initial states have been found using the above procedure we need to apply the monopole boost operator in order to start the nucleus in the breathing mode. This can be done using the usual boost operator for an isoscalar monopole mode
\begin{eqnarray}
Q_\alpha(r_m,0) = e^{ik r_m^2}Q_\alpha(r_m), \label{eq:boost}
\end{eqnarray}
where $k$ is the adjustable strength.

Once this has been done the $Q_\alpha$'s can be propagated in time. The equally spaced time grid
\begin{eqnarray}
t_n=n\Delta t\text{, } n=1,\hdots,N
\end{eqnarray} 
is used and the same space grid, \bref{TIHFspaceGrid}, as the stationary problem. The Crank-Nicholson method is then used for the time discretization of the time-dependent Hartree-Fock equation:
\begin{eqnarray}
&&\brac{\hat{I}+\frac{i\Delta t}{2}\hat{H}(r_m,t_{n-\half})}Q_\alpha(r,t_n) \nonumber \\
&&\qquad= \brac{\hat{I}-\frac{i\Delta t}{2}\hat{H}(r_m,t_{n-\half})}Q_\alpha(r,t_{n-1}) \qquad \nonumber \\
&&\qquad\qquad+ \mathcal{O}(\Delta r^2,\Delta t^2) \label{TDHFCrank}
\end{eqnarray}
We choose the Crank-Nicholson method because it has properties that are useful for this type of calculation: it is unconditionally stable; and it maintains norm. However being an implicit method it also yields the Hamiltonian evaluated at a half time-step and so through the potential term the density evaluated at the half timestep. This means our resulting equations are not a system of linear equations. To get around this problem we use an explicit method, which is calculated after each propagation in time to yield the wavefunctions needed to calculate the half-time-step density. We use a method based on the evolution operator:
\begin{eqnarray} 
Q(r_m,t_{n+\half}) &=& \exp\brac{-\frac{i\Delta t}{2}\hat{H}(r_m,t_n)}Q(r,t_n) \label{intStepEvo}\\
&=& \sum_{j=0}^{j_{max}} \brac{-\frac{i\Delta t}{2}\hat{H}(r_m,t_n)}^n Q(r_m,t_n) \\ &&+ \mathcal{O}(\Delta r^2,\Delta t^{j_{max}})
\end{eqnarray}
requiring knowledge of the Hamiltonian only at the current time-step.

Once equation \bref{TDHFCrank} has been discretized in space using central differences and the grid \bref{TIHFspaceGrid} it is a tridiagonal matrix equation, again solved with LAPACK routines to get from one time to the next. 

However, the last row in the matrix contains an unknown $Q(r_{M+1},t_n)$ for $n>0$. This has to be specified with the boundary condition which we know at infinity, but we require a boundary condition at $r=(M+1)\Delta r$. We could use the same reasoning as the stationary case, that we can find a point at which the wavefunction will be zero and apply the boundary there. We also know however that this system has a probability of particle emission, which manifests itself in the calculations as a thin non-zero tail travelling away from the central mass near the origin. This means as time passes the point at which the wavefunction is zero gets increasingly further away. This corresponds to longer calculation times which can be prohibitive. Hence we seek an absorbing boundary condition to give the value of $Q(r_{M+1},t)$.

\section{\label{sec:probExt}Problem in the Exterior} 

\subsection{Splitting the Domain}

We start by splitting the domain into two regions: an interior in which we choose to contain all the nuclear dynamics; and an exterior where we assume only the long ranged components are of significance, in this case just the centrifugal barrier. Given the partial differential equation for a single particle state in coordinate space:
\begin{eqnarray}
i\pderiv{}{t}Q_l(r,t) = \brac{-\half\pderivtwo{}{r} + V(r,t)}Q_l(r,t), \label{exteriorDE}
\end{eqnarray} 
with boundary conditions:
\begin{eqnarray}
Q_l(0,t)=0, \\
\lim_{r\to\infty}Q_l(r,t)=0. \label{exteriorBC2}
\end{eqnarray}
We can mathematically describe the splitting with the potential term:
\begin{eqnarray}
V(r,t) \equiv V_{short}(r,t) + V_{long}(r),
\end{eqnarray}
where we define:
\begin{eqnarray}
V_{short}(r,t)= 0 &\text{ for }& r \ge R, \label{VintConditions}\\
V_{long}(r)=\frac{l(l+1)}{2r^2} &\text{ for }& r \ge 0 . \label{VextConditions}
\end{eqnarray}
The problem has now been split into where the internal potential is present and where it is not. The parameter $R$ is commonly called the artificial boundary and has to be chosen so equations \bref{VintConditions} and \bref{VextConditions} are satisfied. We also assume that the initial wave function is zero outside the artificial boundary:
\begin{eqnarray}
Q_l(r,0)=0 \text{ for } r \ge R . \label{WFintConditions} 
\end{eqnarray}
This is not overly restrictive and consistent with our choice for the solution of the static Hartree-Fock equations.

\subsection{Deriving the Absorbing Boundary Conditions}
We have now all the assumptions needed to construct the absorbing boundary condition. There are various ways of doing this and a Green's function approached has already been described by Heinen and Kull in \cite{symABC,springerlink:10.1134/S1054660X10050063} for this problem. We proceed differently, however, by describing a derivation using a Laplace transform method.

We start by recalling the definitions of the Laplace transform\cite[Chapter 29]{AbraMathFunc} in time, $\hat{f}(s)$, of a function, $f(t)$, as: 
\begin{eqnarray} 
\hat{f}(s) = \int_{0}^{\infty}f(t)e^{-st}dt, \label{laplaceDef}
\end{eqnarray}
and the inversion formula, known as the Bromwich integral\cite{transMethDuffy}: 
\begin{eqnarray}
f(t) = \frac{1}{2\pi i}\int_{c-i\infty}^{c+i\infty}\hat{f}(s)e^{st}ds. \label{inversionDef}
\end{eqnarray}
Combining equations \bref{exteriorDE} and \bref{VextConditions} for $r\ge R$ we have:
\begin{eqnarray}
i\pderiv{}{t}Q_l(r,t) = \brac{-\half\pderivtwo{}{r} + \frac{l(l+1)}{2r^2}}Q_l(r,t), \label{exteriorDE2}
\end{eqnarray}
Multiplying by $e^{-st}$ and integrating time from $0$ to $\infty$ allows us to use equation \bref{laplaceDef} to get the ordinary differential equation: 
\begin{eqnarray}
\half\pderivtwo{\hat{Q_l}(r,s)}{r} +\brac{is - \frac{l(l+1)}{2r^2} }\hat{Q_l}(r,s) = 0.
\end{eqnarray}
The substitution $Q_l(\rho,s)=\rho h_l(\rho,s)$ where $\rho = kr$ and $k=\sqrt{2is}$, yields the following equation for $h_l(\rho,s)$:
\begin{eqnarray}
r^2\pderivtwo{h_l}{\rho} +2r\pderiv{h_l}{\rho} +\brac{r^2-l(l+1)}h_l = 0,
\end{eqnarray}
where the square root is assumed to be on the branch resulting in a positive real part. As $l\in\mathbb{N}_0$ we can see that this equation has spherical Bessel functions as solutions \cite[Chapter 10]{AbraMathFunc} of which there are various satisfactory pairs. We choose the particular solutions as the spherical Bessel functions of the third kind, also known as spherical Hankel functions. Any pair of solutions can be used to give the same end result once the boundary condition are applied. However this pair simplifies the consequent derivations.

Taking the Hankel function solutions,  we can write $\hat{Q_l}$ as:
\begin{eqnarray}
\hat{Q_l}(r,s) = \left. A(s)\rho h_l^{(1)}(\rho)+B(s)\rho h_l^{(2)}(\rho)\right|_{\rho=kr}.
\end{eqnarray}
Only the boundary condition \bref{exteriorBC2} is relevant here, to be precise its Laplace transform, as $r\ge R$ and may be applied by the use of the following limiting forms for $z\to\infty$:
\begin{eqnarray}
&&h_l^{(1)}(z) \sim i^{-l-1}z^{-1}e^{iz}, \label{h1limiting}\\
&&h_l^{(2)}(z) \sim i^{l+1}z^{-1}e^{-iz}.
\end{eqnarray}
Assuming $c>0$ in the Bromwich integral \bref{inversionDef} allows us to say that $y>0$ where $k=\sqrt{2is}=x+iy$, along the integration path. So by the limiting form of $\hat{Q}(r,s)$ as $r\to\infty$:
\begin{eqnarray}
\hat{Q_l}(r,s) \sim A(s)i^{-l-1}e^{(ix-y)r}+B(s)i^{l+1}e^{(y-ix)r},
\end{eqnarray}
we must have $B(s)=0$.

$\hat{Q}(r,t)$ and its $r$ derivative can now be written as:
\begin{eqnarray}
\hat{Q_l}(r,s) = \left. A(s)\rho h_l^{(1)}(\rho)\right|_{\rho=kr}, \\
\pderiv{\hat{Q}_l(r,s)}{r} = \left. A(s)k\pderiv{}{p}\brac{\rho h_l^{(1)}(\rho)}\right|_{\rho=kr}.
\end{eqnarray}
Division of these two equations and evaluating on the artificial boundary yields the Laplace transform of the absorbing boundary condition:
\begin{eqnarray}
\hat{Q}_l(R,s) = \brac{\left.\frac{1}{k}\frac{ \rho h_l^{(1)}(\rho)}{ \pderiv{}{p}\brac{\rho h_l^{(1)}(\rho)}}\right|_{\rho=kr}}\pderiv{\hat{Q_l}(R,s)}{r}.
\end{eqnarray}
Use of the convolution theorem for Laplace transforms gives us the absorbing boundary condition:
\begin{eqnarray}
Q_l(R,t) = \int_0^{t} G_l(R,\tau)\pderiv{Q_l(R,t-\tau)}{r}  \, d\tau, \label{ABCresult1}
\end{eqnarray}
where we define:
\begin{eqnarray}
\hat{G}_l(R,s) \equiv \left.\frac{1}{k}\frac{ \rho h_l^{(1)}(\rho)}{ \pderiv{}{p}\brac{\rho h_l^{(1)}(\rho)}}\right|_{\rho=kr}. \label{laplaceABC1}
\end{eqnarray}
$\hat{G}(R,s)$ being the Laplace transform of $G(R,\tau)$, which can be simplified by the recurrence relation:
\begin{eqnarray}
\deriv{\hl(z)}{z} = \frac{n}{z}\hl(z) - h_{l+1}^{(1)}(z),
\end{eqnarray}
to:
\begin{eqnarray}
\hat{G}_l(r,s) \equiv \left.\frac{1}{k}\frac{ \rho h_l^{(1)}(\rho)}{ (l+1)h^{(1)}_l(\rho)-ph^{(1)}_{l+1}(\rho) }\right|_{\rho=kr}.
\end{eqnarray}

\subsection{Calculation of the kernel $G(r,t)$}
Our final task, before discretization, is to calculate the inverse Laplace transform above. This is done by using a series expansion\cite[p439]{AbraMathFunc} for $h_l^{(1)}(z)$:
\begin{eqnarray}
&&h_l^{(1)} = i^{-l-1}z^{-1}e^{iz}\sum^l_0(l+\half,k)(-2iz)^{-k},
\end{eqnarray}
where:
\begin{eqnarray}
&&(l+\half,k) = \frac{(l+v)!}{v!(l-v)!}.
\end{eqnarray}
After manipulation and simplification we gain the rational function in $k$:
\begin{eqnarray}
\hat{G}_l(R,s) = \frac{ -i\sum_{v=0}^l \sbrac{\frac{(l+\half,v)}{(l+\thrhalf,0)(-2iR)^{v}}} k^{l-v} } { k^{l+1} + \sum_{v=0}^l \sbrac{ \frac{ (l+\thrhalf,v+1) - 2(l+1)(l+\half,v)}{(l+\thrhalf,0)(-2iR)^{v+1}} }k^{l-v}   }. \label{eqn:SSReadyForPF}
\end{eqnarray}
This can be expanded in partial fractions:
\begin{eqnarray}
\hat{G}_l(R,s) &=& \sum^{l+1}_{j=1} \frac{\alpha_j}{k-k_j} \label{PFresult} \\
&=& \sum^{l+1}_{j=1} \frac{\frac{\alpha_j}{\sqrt{2i}}}{\sqrt{s}-\frac{k_j}{\sqrt{2i}}}, \label{PFresult2}
\end{eqnarray}
where the $k_j$ are the roots of the polynomial in the denominator of \bref{eqn:SSReadyForPF} and $a_j$ are the pole strengths. In practice we calculate the roots and strengths for each $l$ with Maple.

The inversion of \bref{PFresult2} is performed just by applying the well known result from tables\cite{AbraMathFunc,intTransErdelyi}:
\begin{eqnarray}
\mathcal{L}^{-1}\left\{ \frac{1}{\sqrt{s}+a}\right\} = \frac{1}{\sqrt{\pi t}} - a\mathrm{w}(ia\sqrt{t}),
\end{eqnarray}
rather than contour integration of the Bromwich integral \bref{inversionDef}. Here $\mathrm{w}(z)=e^{-z^2}\mathrm{erfc}(-iz)$ is the Faddeeva function, which can be calculated with an implementation of reference \cite{Poppe:1990:MEC:77626.77629}. $G(R,s)$ can now be written as:
\begin{eqnarray}
G_l(R,\tau) = \sum^{l+1}_{j=1} \sbrac{\frac{\alpha_j}{\sqrt{2\pi it}}-\half i\alpha_j k_j\mathrm{w}\brac{z_j}},
\end{eqnarray}
where $z_j = -k_j\sqrt{\frac{i\tau}{2}}$. Simplification of the above can be made by using the limiting form \bref{h1limiting} in equation \bref{laplaceABC1} and comparing to \bref{PFresult} in the limit $k\to\infty$:   
\begin{eqnarray}
0=\lim_{k\to\infty} \brac{k\hat{G}_l(r,s) - k\hat{G}_l(r,s)}\qquad\qquad\qquad\quad \\
= \lim_{k\to\infty} \brac{\left. \rho\frac{i^{-l-1}e^{i\rho}}{\pderiv{}{\rho}(i^{-l-1}e^{i\rho})}\right|_{\rho=kr}  -  \sum_{j=1}^{l+1} \alpha_j\frac{k}{k-k_j}},
\end{eqnarray}
the differentiation of the limiting form is allowed as the functions $\hl(z)$ are analytic. The limit can be performed to give:
\begin{eqnarray}
\sum_{j=1}^{l+1} \alpha_j = -i,
\end{eqnarray}
which allows us to write our final form of the kernel $G$ as:
\begin{eqnarray}
G_l(R,\tau) = \frac{-i}{\sqrt{2\pi i \tau}}-\frac{i}{2}\sum^{l+1}_{j=1} \alpha_j k_j\mathrm{w}\brac{z_j}. \label{ABCkernel1}
\end{eqnarray}
An interesting and reassuring feature of this boundary condition is that for $l=0$ where equation \bref{exteriorDE} reduces to the free one dimensional Schrödinger equation, we have the values $a_1=-i$ and $k_1=0$.  Using these values we gain the absorbing boundary condition for the free one dimensional Schr\"odinger equation as found in \cite{symABCprev}.

\section{\label{sec:boundDisc}Boundary Discretization}

\subsection{Removing the Singularity}
Equations \bref{ABCresult1} and \bref{ABCkernel1} will now be discretized on the grid for use in the Crank-Nicholson scheme. Inspecting equation \bref{ABCkernel1} we see that it has a square root singularity at $\tau=0$ and is not ideal for numerical integration. So integration by-parts is done on the first term to give:
\begin{eqnarray}
G_l(R,\tau) = \sqrt{\frac{2i\tau}{\pi}}\pderiv{}{\tau} - \frac{i}{2}\sum^{l+1}_{j=1} \alpha_j k_j\mathrm{w}\brac{z_j}. \label{ABCkernel2}
\end{eqnarray}
Our function is now continuous at $\tau=0$ and although its derivatives are not it is better suited to the numerical integration. Note that $G_l(R,\tau)$ is now an operator. Defining a function $u^{(l)}(R,\tau)$ allows for a more compact expression:
\begin{eqnarray}
G_l(R,\tau) = \sqrt{\frac{2i\tau}{\pi}}\pderiv{}{\tau} + u^{(l)}(R,\tau). \label{ABCkernel3} \\ u^{(l)}(R,\tau) = - \frac{i}{2}\sum^{l+1}_{j=1} \alpha_j k_j\mathrm{w}\brac{z_j}
\end{eqnarray}

\subsection{Time Discretization}

We first form a semi-discrete equation on the grid $t_n = n\Delta t$ with $t=t_N$ and $\tau_n=t_n$. By using the extended midpoint rule:
\begin{eqnarray}
\int_0^t f(\tau) \, d\tau = \Delta t \sum_{n=0}^{N-1} f\brac{t_{n+\half}} + \mathcal{O}(\Delta t^2)
\end{eqnarray}
to evaluate the integral and the difference formulas:
\begin{eqnarray}
f(r,t_{n-\half}) = \frac{f(r,t_{n})+f(r,t_{n-1})}{2} + \mathcal{O}(\Delta t^2)\\
\pderiv{f(r,t_{n-\half}) }{t} = \frac{f(r,t_{n})-f(r,t_{n-1})}{\Delta t} + \mathcal{O}(\Delta t^2)
\end{eqnarray}
for functions evaluated at a half time step gives the following semi-discrete equation:
\begin{eqnarray*}
&&Q_l(R,t_N) +\brac{\sqrt{\frac{2it_\half}{\pi}} - \frac{\Delta t}{2}u_l(R,t_\half)}\deriv{Q_l(R,t_N)}{r} \\
&=& \brac{\sqrt{\frac{2it_\half}{\pi}} + \frac{\Delta t}{2}u_l(R,t_\half)}\deriv{Q_l(R,t_{N-1})}{r}
\end{eqnarray*}
\begin{eqnarray*}
\quad- \sum_{n=1}^{N-1} \brac{\sqrt{\frac{2it_{n+\half}}{\pi}} - \frac{\Delta t}{2}u_l(R,t_{n+\half})}\deriv{Q_l(R,t_{N-n})}{r}\quad&& \nonumber \\
+ \sum_{n=1}^{N-1} \brac{\sqrt{\frac{2it_{n+\half}}{\pi}} + \frac{\Delta t}{2}u_l(R,t_{n+\half})}\deriv{Q_l(R,t_{N-n-1})}{r}&& \\
+\mathcal{O}(\Delta t^2)\qquad&&
\end{eqnarray*}

\subsection{Space Discretization} 
For the space discretization we choose the artificial boundary at $R=r_{M-\half}$ between the penultimate and final spatial grid-points. The following difference formulas are used:
\begin{eqnarray}
f(r_{M-\half},t) = \frac{f(r_M,t)+f(r_{M-1},t)}{2} + \mathcal{O}(\Delta r^2)\\
\pderiv{f(r_{M-\half},t) }{t} = \frac{f(r_{M},t)-f(r_{M-1},t)}{\Delta t} + \mathcal{O}(\Delta t^2)
\end{eqnarray}
at the points between the spatial grid. This yields the fully discetized absorbing boundary condition:
\begin{eqnarray}
&&\!\!\!\!\!\!\!\!\brac{1-B^{(M,0)}_l}Q_l(r_M,t_N) + \brac{1+B^{(M,0)}_l}Q_l(r_{M-1},t_N) \nonumber \\
&=&C^{(M,0)}_l\brac{Q_l(r_{M-1},t_{N-1})-Q_l(r_M,t_{N-1}) \phantom{\frac{}{}} } \qquad\qquad\nonumber  \\
&+& \sum_{n=1}^{N-1}B^{(M,n)}_l\brac{Q_l(r_M,t_{N-n}) - Q_l(r_{M-1},t_{N-n}) \phantom{\frac{}{}} } \qquad \nonumber\\
&+& \sum_{n=1}^{N-1}C^{(M,n)}_l\brac{Q_l(r_{M-1},t_{N-n-1}) - Q_l(r_M,t_{N-n-1}) \phantom{\frac{}{}} }\nonumber\\
&+& \mathcal{O}(\Delta r^2,\Delta t^2).\label{discreteABC}
\end{eqnarray}
Where:
\begin{eqnarray*}
&&A = \frac{-2}{\Delta r}\sqrt{\frac{i\Delta t}{\pi}}, \\
&&B^{(M,n)}_l = A\sqrt{2n+1} + \frac{\Delta t}{\Delta r}u_l(r_{M-\half},t_{n+\half}), \\
&&C^{(M,n)}_l = A\sqrt{2n+1} - \frac{\Delta t}{\Delta r}u_l(r_{M-\half},t_{n+\half}).
\end{eqnarray*}
Within the implementation, equation \bref{discreteABC} replaces the last row of the matrix described in section \bref{section_IntDiscTDT}.

\section{\label{sec:ResultsAndTesting1}Results and Testing: Absorbing Boundary Effectiveness}

Before calculating the giant resonances, the implementation of the absorbing boundary is tested in a simplified case, without any potential, beyond that coming from the centrifugal term.  We apply the absorbing boundaries to a partial differential equation of the form \bref{exteriorDE}. This is to show the validity of the implementation and to demonstrate its performance. The solution to the following partial differential equation is found:
\begin{eqnarray}
i\pderiv{Q_l}{t} = \half\pderivtwo{Q_l(r,t)}{r} + \frac{l(l+1)}{2r^2}Q_l(r,t), \label{results:DE1}
\end{eqnarray}
\begin{eqnarray}
&Q_l(r,0)= Are^{-(r-5)^2} ,& \\
&Q_l(0,t)=0 \text{,\quad} \lim_{r\to\infty} Q_l(r,t) = 0, &
\end{eqnarray} 
for $l=0,1,2$. Although calculations can be done for any angular momentum these are the only values required for the Hartree-Fock calculations shown later. $A$ is chosen to normalise $Q_l(r,0)$ and is calculated with Simpson's rule.

 Physically the equation corresponds to the evolution of a free particle which initially is a shell surrounding the origin. Although this sort of system provides no particular physical insights, it does allow us to make quick and simple calculations which are suitable for testing the validity of the method.

We use the same time and space discretization as described in section \bref{sec:intDisc} to discretise equation \bref{results:DE1}. The intermediate step \bref{intStepEvo} is not needed here, as the equation is linear.

Our results will show comparisons between a calculation done with absorbing boundaries at $r=10$ and one with reflecting boundaries at a radius chosen so reflection does not occur, which will be specified for each test.

For this simplified case, we take $\hbar=m=1$. 

\subsection{Densities}

\begin{figure}[!htb]
%
%
%
%
%
\centerline{\includegraphics[scale=1.35]{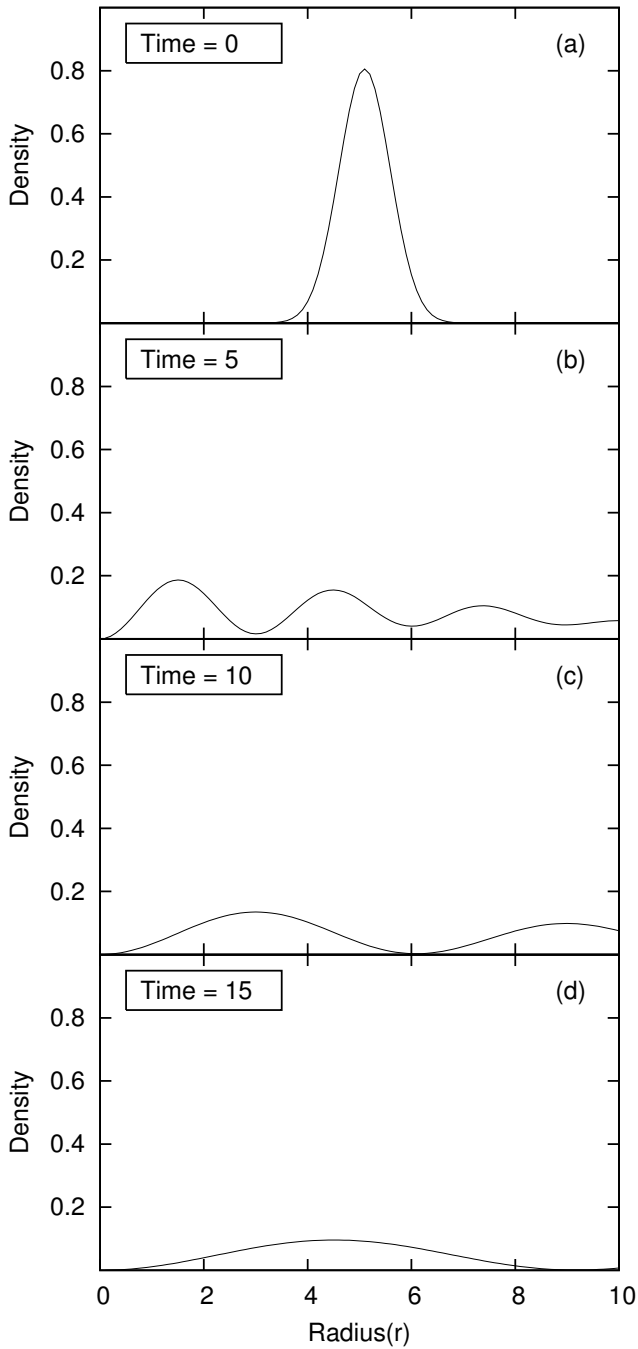}}
\caption{These figures show wavefunctions, of angular momentum $l=0$ changing in time with a percentage leaving the interval of interest. The calculations are done with a reflecting boundary at $r=100$ and have grid spacings of $\Delta x = \Delta t = 0.1$. From top to bottom the graphs show the evolution of the wavefunctions at times 0,5,10 and 15.}
\label{result1}
\end{figure}

To show how the solutions to equation \bref{results:DE1} evolve through time the probability densities are presented. These are gained from calculating the wavefunction through time with  a reflecting boundary at $r=100$. In the time interval chosen, $[0,15]$, reflection does not occur. Figure \ref{result1} shows us the densities through time for each angular momentum. Only the interval $[0,10]$ is plotted as this is where we place the test absorbing boundary. The results are calculated with grid spacings $\Delta x=\Delta t=0.1$.

In each case we see the bulk of the density begins centred at $r=5$.  As it the system evolves, the wavepacket spreads out, and interferes with itself as it reaches the origin.

\subsection{Radial Comparison of Wavefunction}

\begin{figure}[!htb]
%
%

%
\centerline{\includegraphics[scale=1.35]{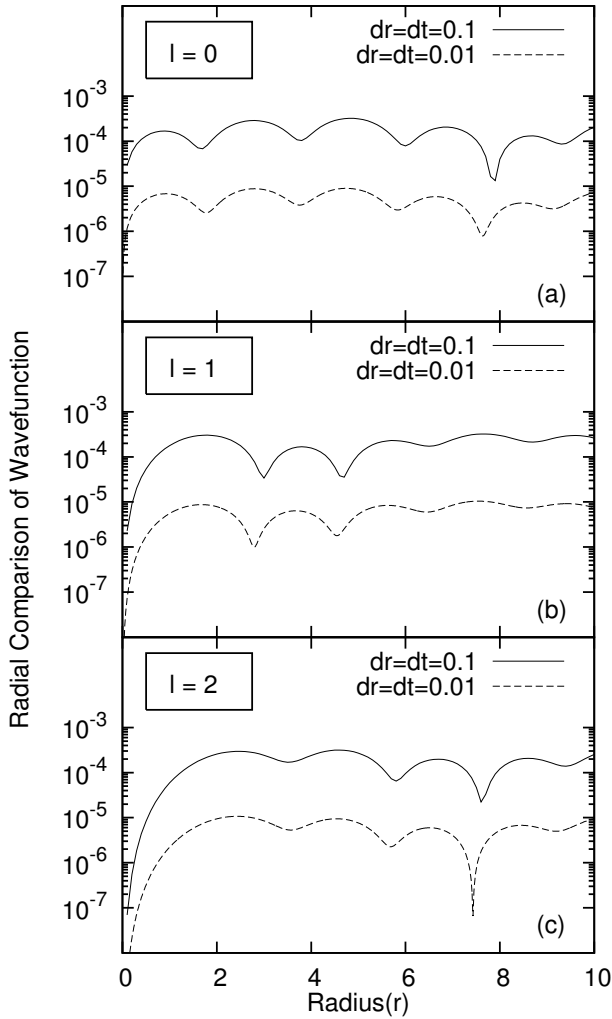}}
\caption{The figures shows a comparison of the radial component of the wavefunctions at the final time $15$, for angular momenta $l=0,1,2$, calculated with each technique. The value in equation \bref{test2eqn} is plotted against the radius.}
\label{results2} 
\end{figure}

We now go on to see how the absorbing boundary performs. We plot:
\begin{eqnarray}
|Q^{(Ref)}_l(r,t)-Q^{(ABC)}_l(r,t)|
\label{test2eqn}
\end{eqnarray}
at $t=15$, where $Q^{(Ref)}_l$ and $Q^{(ABC)}_l$ are the calculations with reflecting and absorbing boundaries respectively. This is to see how any error from the absorbing boundary effects the interior points. Figure \ref{results2} shows the result for each angular momentum with two different grid spacings. Again the reflecting boundaries are chosen to be at $r=100$.

We see that in all cases the error has remained small throughout the interior, for the $dx=dt=0.1$ case bounded by $10^{-3}$ and for $dx=dt=0.01$ bounded by $10^{-5}$. This is within the $\mathcal{O}(\Delta r^2,\Delta t^2)$ expected from the discretisation.

\subsection{Temporal Comparison of Probability}

We now test the how the error evolves through time. This is done by calculating the probability of finding the particle inside the interval over time, mathematically the following is calculated:
\begin{eqnarray}
P(t) = \int_0^{10} |Q_l(r,t)|^2 \, dr
\label{test3eqn}
\end{eqnarray}
with reflecting and absorbing boundaries and the absolute value of the difference taken.

For this test we increase the time interval to $[0,50]$ and move the reflecting boundary to $r=200$. In each case more than $90\%$ of the wavefunction has left the interval, specifically the probabilities inside the interval are $8.57E-002$, $6.36E-003$ and $2.03E-004$ for $l=0,1,2$ respectively at the end of the calculation.

\begin{figure}[!htb]
%
%
%
%


\centerline{\includegraphics[scale=1.55]{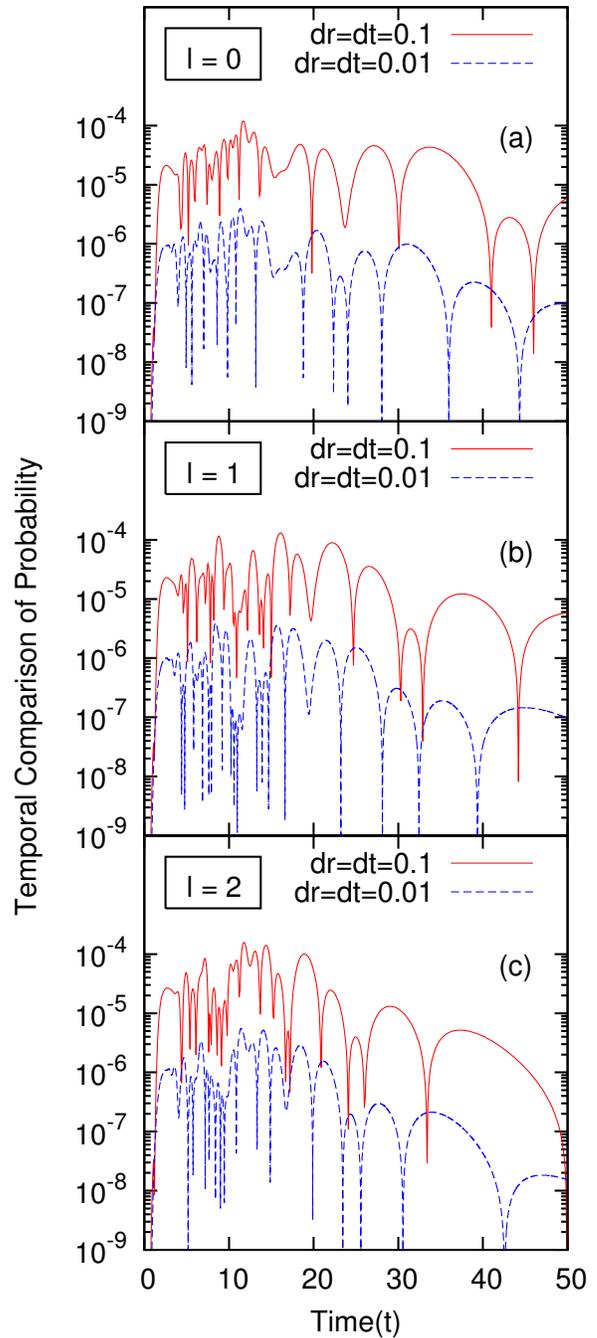}}
\caption{(Color online) These plots show how the error in the probability from the absorbing boundaries changes through time. Equation \bref{test3eqn} is calculated with reflecting and absorbing boundaries and the absolute value of there difference taken, though time and plotted.}
\label{results3}
\end{figure}

Figure \ref{results3} shows the results for each angular momenta and different grid spacings.

We see that in time also the error remains bounded. From the plots it appears the bound on the error is proportional to the grid spacings.

These results are satisfactory and so now with confidence in the previous work we go on to the Hartree-Fock calculations.

\section{\label{sec:ResultsAndTesting2}Results and Testing: Hartree-Fock Resonances in the Linear Regime}
Results from the implementation of the discretised Hartree-Fock system, as described in sections \ref{sec:intDisc} and \ref{sec:boundDisc}, are now shown. We first present the variation of the root mean square radius over time for $\nucl{4}{2}{He}$, $\nucl{16}{8}{O}$ and $\nucl{40}{20}{Ca}$. For each nuclei the following is shown:
\begin{enumerate}[(a)]
\item A calculation performed with reflecting boundaries at $1500$ fm. This is the result expected from a continuum calculation because the boundary is far enough away so as to avoid reflection. This is plotted from $0$ to $500 \text{ fm c}^{-1}$ to show the main features occurring at the beginning of the resonance.

\item The result of using reflecting boundaries at 30 fm. This is to show the effect the absorbing boundaries are having. Again this is plotted from $0$ to $500 \text{ fm c}^{-1}$. 

\item The difference between the expected result in (a) and a calculation with absorbing boundaries at $30$ fm. This is plotted for the entire 0 to $3000 \text{ fm c}^{-1}$ time range. This difference is an error due to the discretization of the absorbing boundaries and so we consider a upper bound for this value of $\mathcal{O}(\Delta r^2)$ acceptable.
\end{enumerate}
For each nucleus has a group of three figures are shown which are labelled according to the above. We also show the time each calculation takes to evaluate the efficiency of the absorbing bounds.

Grid spacings of $\Delta r = 0.1\text{ fm}$ and $\Delta t = 0.1\text{ fm c}^{-1}$ are used and all calculation are evolved from 0 to 3000 fm c$^{-1}$.

\subsubsection{Helium-4}

\begin{figure}
	\centerline{\includegraphics[scale=1.35]{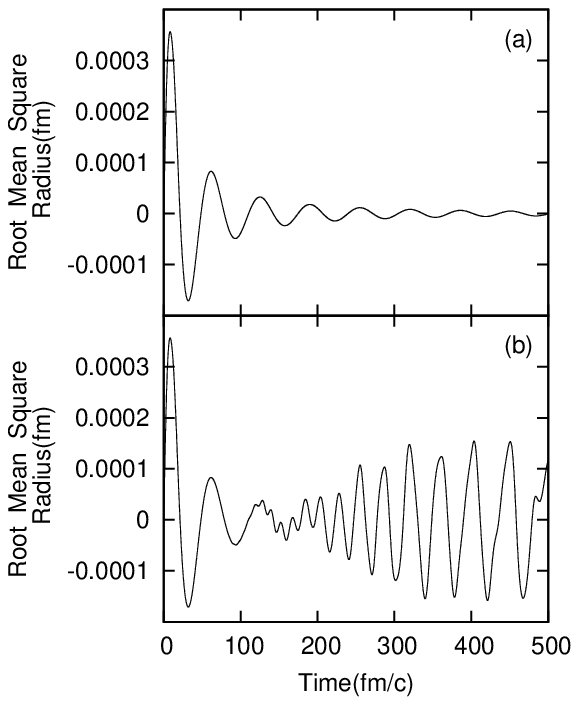}}
	\centerline{\includegraphics[scale=1.35]{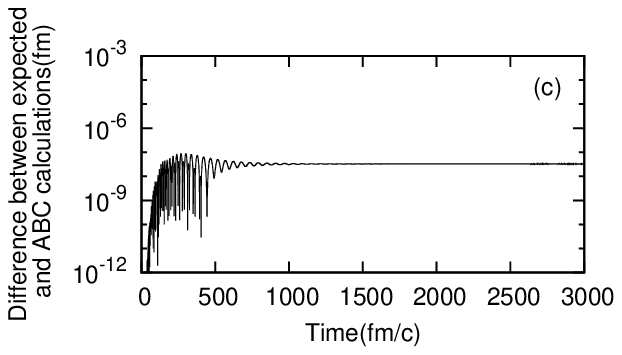}}
	\caption{The time evolution of the monopole moment in Helium-4, showing (a) the continuum result, (b) for comparison, the result of a reflecting boundary wall and (c) the absolute value of the difference between the monopole moments when calculated using a absorbing boundary and using a far reflecting wall, over time.}	
\label{test_He}
\end{figure}
From figure (\ref{test_He}a) we can see that the resonance for $\nucl{4}{2}{He}$ has a simple damped oscillatory motion, the radius of the nuclei repeatedly increasing and decreasing clearly demonstrating the breathing mode. Figure (\ref{test_He}c) shows us that the absorbing boundary provide us with a reasonable discrepancy from the expected result being bounded by $10^{-7}$, well below the $\mathcal{O}(0.1^2)$ discretization error. Finally by comparing (\ref{test_He}a) and (\ref{test_He}b) the effect of the reflected flux can clearly be seen, which is the source of discretisation artefacts in the strength functions \cite{Stevenson2007}.


\subsubsection{Oxygen-16}

\begin{figure}
	\centerline{\includegraphics[scale=1.35]{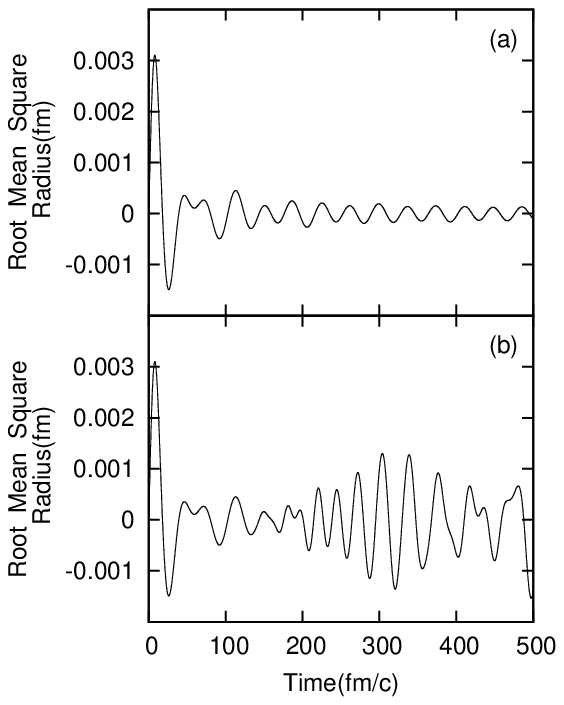}}
	\centerline{\includegraphics[scale=1.35]{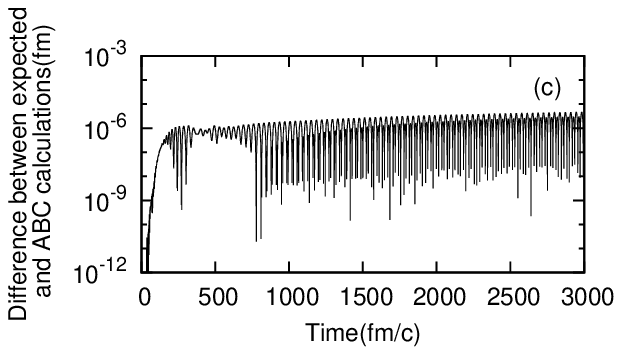}}
	\caption{The time evolution of the monopole moment in Oxygen-16, showing (a) the continuum result, (b) for comparison, the result of a reflecting boundary wall and (c) the absolute value of the difference between the monopole moments when calculated using a absorbing boundary and using a far reflecting wall, over time.}	
\label{test_O}
\end{figure}

The top panel of figure (\ref{test_O}a) shows a more complicated motion of the nucleus this time, which does not look like a single damped mode. This is due to the multiple single-particle states present, known as Landau fragmentation. The absolute error as shown in figure (\ref{test_O}c) is bounded by a larger number than helium, but again within the acceptable range. 

\subsubsection{Calcium-40}

\begin{figure}
\centerline{\includegraphics[scale=1.35]{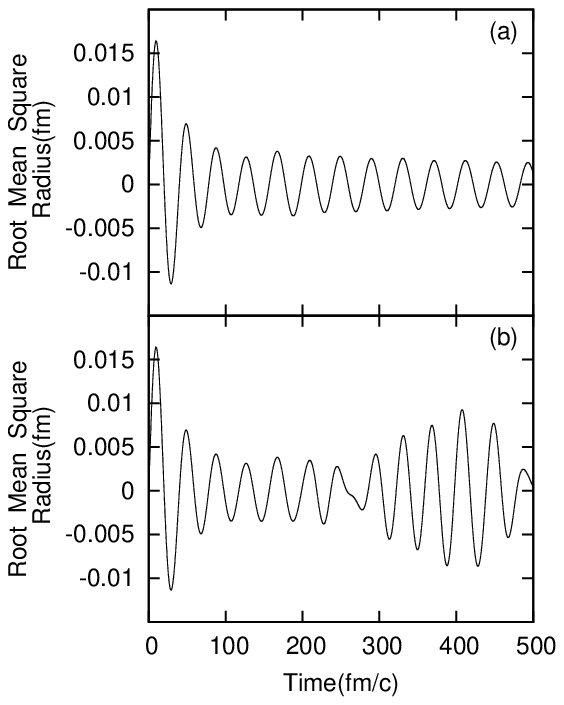}}
\centerline{\includegraphics[scale=1.35]{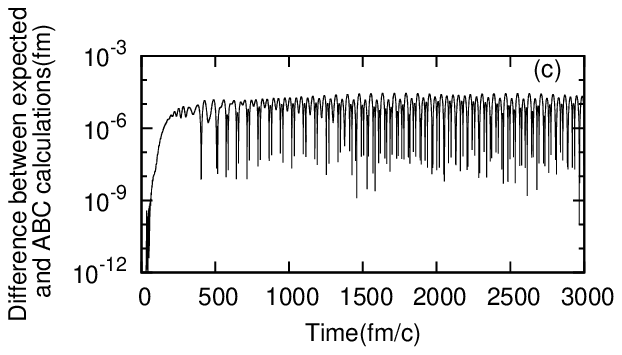}}	
	\caption{The time evolution of the monopole moment in Calcium-40, showing (a) the continuum result, (b) for comparison, the result of a reflecting boundary wall and (c) the absolute value of the difference between the monopole moments when calculated using a absorbing boundary and using a far reflecting wall, over time.}	
\label{test_Ca}
\end{figure}

The results for calcium again show a damped oscillation, as expected, though a long-lived resonant component is excited too, which the reflecting boundaries obviously cannot reproduce for long times.  The errors are somewhat larger than the helium or oxygen cases but still acceptable.

\subsection{Timing}
As an guide, we present a table of timing results for the Oxygen calculations in Table \ref{tab:timing}.

\begin{table}[tbh]
\begin{tabular}{|l|c|c|}
\hline
	Boundary Type & R(fm) & Calculation Time (s)\\
	\hline
	Reflecting & 1500	& 2378 \\
	Reflecting & 30		& 58  \\
	Absorbing  & 30		& 144\\
\hline
\end{tabular}
\caption{Calculation times for the large box continuum calculation with reflecting bounds, a small-box calculation with spurious reflections and a small-box calculation with absorbing boundaries.\label{tab:timing}}
\end{table}

\begin{figure}[htp]
\centering
\includegraphics[scale=1.35]{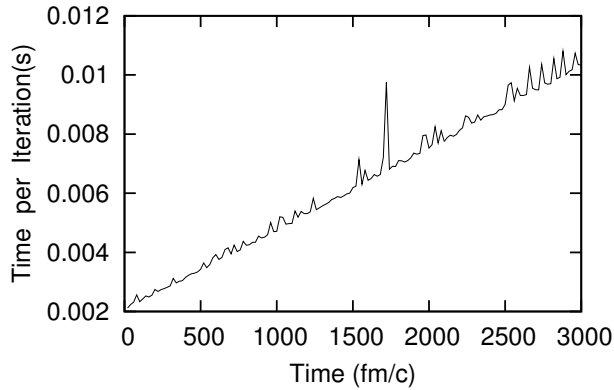}
\caption{A plot showing the expense of each iteration in a calculation of oxygen-16. It clearly show the non-locally of the absorbing boundary increasing the calculation time for iteration the further the calculation progresses.}
\label{fig:itertime}
\end{figure}

The results show that the absorbing boundaries are considerably more expensive than reflecting boundaries, but less so than using a large box with simple boundary conditions.  It is interesting also to examine the time taken to each iteration. Figure \bref{fig:itertime} shows a plot of the time to compute each iteration, as a running average over 20 iterations to somewhat smooth out the effect of computer load.
This shows the steady increase in expense to calculate a iteration as the calculation progresses and is due to the non-locality in time of the absorbing boundary condition.

\subsection{Strength Functions}

\begin{figure}[!htb]
	\includegraphics[scale=1.3]{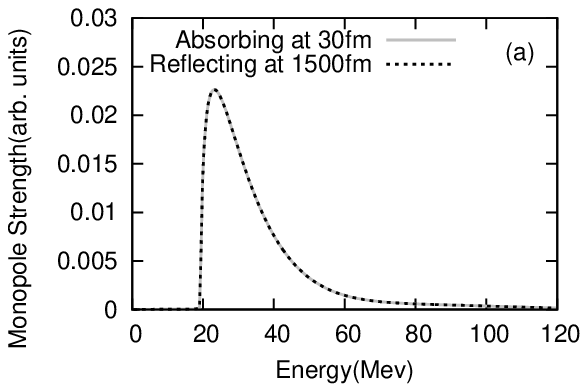}
	\includegraphics[scale=1.3]{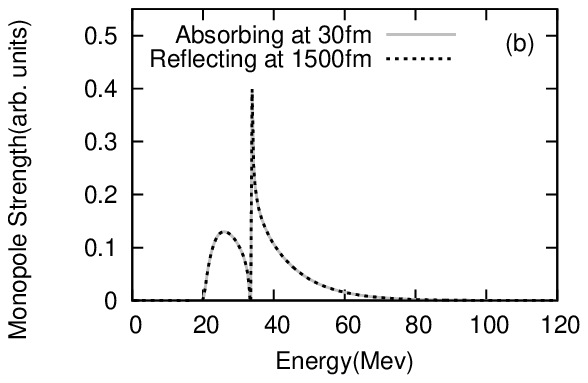}
	\includegraphics[scale=1.3]{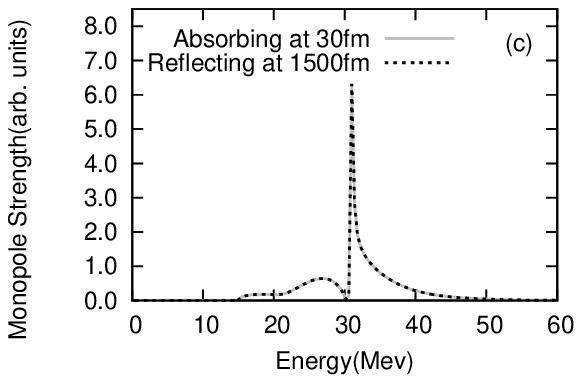}
\caption{Plots showing the effect of using the absorbing boundary condition on the strength functions of various nuclei. Going from top to bottom there is the helium, oxygen and calium strength functions.}
\label{resultsz}
\end{figure}

The strength functions for these calculations are now presented. As these are the calculations required in order to make comparisons to experiment their accurate calculation is critical. We require that the error in the above results do not give noticable artefacts in the strength functions, at least to the level of experimental resolution. Figure \bref{resultsz} shows the calculated strength function from the expected result with that calculated using absorbing boundaries.

We see that both calculations match up well for all the nuclei tested. The figures show the increasing complexity of the nuclear structure, as more features appear in the strength functions.

\section{\label{sec:ResultsAndTesting3}Results and Testing: Non-Linear Regime}
\begin{figure}[tbh]
\includegraphics[scale=1.5]{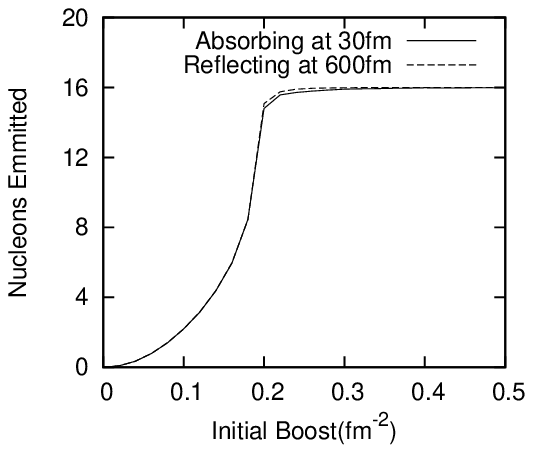}
\caption{\label{fig:emission_summary}
A comparison of the number of particles emitted from the region between 0 and 30fm with absorbing boundaries at 30fm compared with reflecting boundaries at 600fm which are not reached in the time of the calculation.
}
\end{figure}
\begin{figure}[tbh]
\includegraphics[scale=1.5]{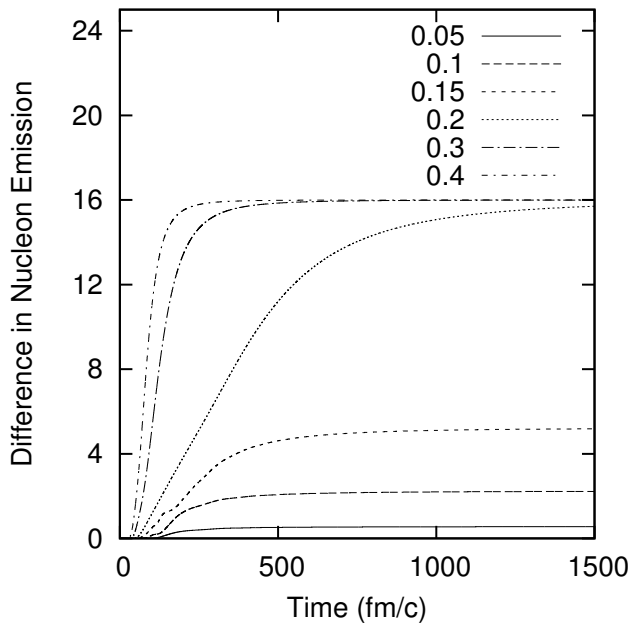}
\caption{\label{fig:particlestime}
The time-dependence of particle emission as a function of boost strength for large-amplitude excitations in $^{16}$O.  The legend indicates the strength $k$ (fm$^{-2}$) of the boost in equation (\ref{eq:boost}).
}
\end{figure}

\begin{figure*}[tb]
  \centerline{\includegraphics[scale=1.12]{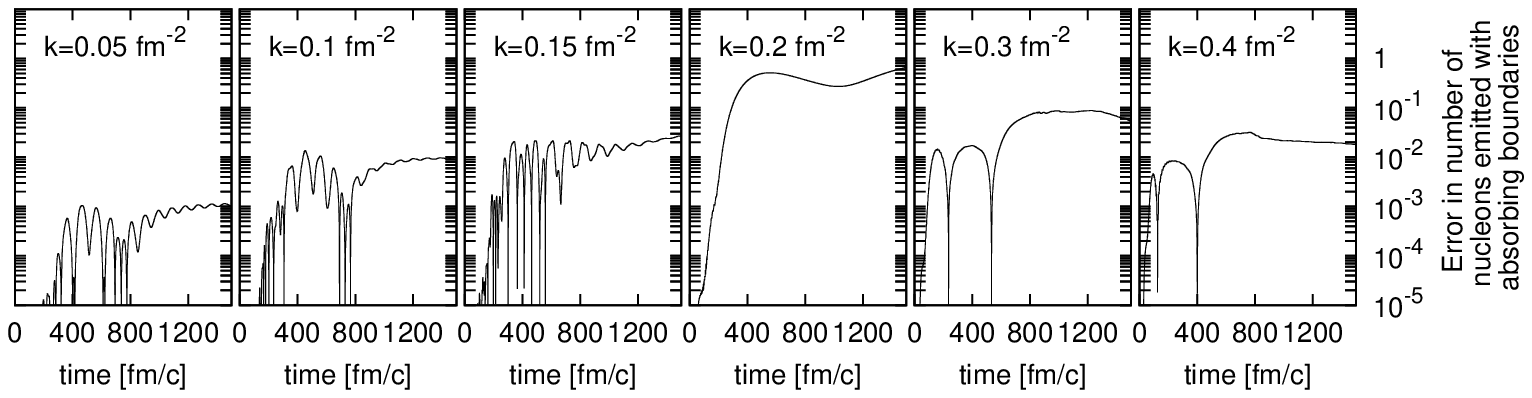}}
\caption{The total error in number of particles emitted by the nucleus as a function of time for increasingly stronger boosts (indicated by the strength $k$ in each panel).  The error is calculated with respect to a calculation withouth aborsbing bounds but in a space so large that the boundaries are not probed.  The boost parameter $k$ is as defined in (\ref{eq:boost}).\label{parterrstime}}
\end{figure*}
As well as testing in the small amplitude linear response regime, of relevance to giant resonances, it is also instructive to examine the larger-amplitude regime, which can be studied in THDF-based techniques \cite{PhysRevC.68.024302,PhysRevC.80.064309,Reinhard2007}, unlike the small-amplitude-limited RPA.  This regime is relevant to the decay of highly excited fragments following e.g. deep inelastic collisions, and significant particle emission may be expected.  Similar situation arise in atomic physics where direct electromagnetic excitation of highly ionizing collective modes is feasible \cite{cluster}.  We use a test case of monopole exciations of $^{16}$O, with increasingly strong boosts (\ref{eq:boost}) such that eventually all particles are lost from the nucleus through large-amplitude excitation.  We note that the computational effort for large amplitude excitations is not different to that for small-amplitude excitations, as the iteration procedure is not changed for larger amplitudes.

Despite the success of the small-amplitude calculations, there is no a priori reason to expect larger amplitude calculations to perform so well, since our absorbing boundaries are predicated on the fact that the only potential active at the boundary is the centrifugal barrier, whereas the nuclear mean-field exists wherever the nucleon wavefunction is finite.  As more particles are emitted, so too the nuclear wavefunction and its associated mean-field are present in the exterior region.  Figure \ref{fig:emission_summary} shows the comparison of the total number of particles emitted (by 1500 fm/c) from a $^{16}$O nucleus between an absorbing boundary calculation, and a reflecting boundary calculation in which the size of the box is so high that the reflecting boundaries are not reached.  The range of boost is sufficiently large to cover the small amplitude limit as well as the regime in which the nucleus is entirely ionized. The two calculations are seen to be close over the entire range, with small differences near the bend as complete ionization occurs.  The time-dependence of the particle emission is shown in Figure \ref{fig:particlestime}, in which the case around the bend is shown to still be changing at the end time of the calculation.

Figure {\ref{parterrstime}} shows the time-dependent error (absorbing bounds compared with large-space reflecting bounds) in the total number of particles emitted for a range of kick size.  This highlights the small differences in Figure \ref{fig:emission_summary} where the errors around $k=0.2$ fm$^{-2}$ are seen to be largest.  In the worst case, this error is noticable, but still rather small.

\section{Perspectives and Conclusion} 

\subsection{Perspectives for more realistic calculations}

Our calculations respresent a step on the way to more realistic calculations of giant resonances within a continuum time-dependent Hartree-Fock framework.  We discuss in this section some perspectives for the possibility of performing more realistic calculations.  Our calculations deliberately considered a simple case, yet within TDHF-based methods, calculations without our form of absorbing bound exist with more relaxed symmetries \cite{PhysRevC.71.064328,brinemg,PhysRevC.71.024301,Umar2005} or with pairing in the BCS or TDHFB framework \cite{PhysRevC.71.064328,PhysRevC.82.034306,PhysRevC.78.044318,PhysRevC.84.051309}.  Our method is extendable in a straightforward way to calculations involving pairing.  The increased expense scales in the same way as discrete calculations with pairing scale with respect to calculations without pairing.  The addition of extra single-particle states to account for the scattering of Cooper pairs will involve extra boundary conditions, but only with a linear scaling with respect to the number of particle states.  On the other hand, increased dimensions will be more costly.  In our case of spherical symmetry in which there is a single boundary point for ~300 interior points, we have a similar time spent on the boundary as the entire internal region. In a three-dimensional calculation, in which the boundary is the surface of volume, the ratio of boundary points to internal points is much higher.  Our technique is thus not currently suitable for a three-dimensional calculation.  However, reasonable scaling could nevertheless be achieved with an expansion of the density in spherical harmonics.  For the purposes of calculating giant resonances of general multipolarity and of deformed nuclei, this would suffice, as only one point per moment of the density would be needed to act as a boundary point, and a typical expansion of a handful of terms would describe a small-amplitude deformation.  A full three-dimensional code would remain required for heavy-ion collisions.

Our immediate aim is to find a suitable way to include the Coulomb potential, which has been ignored here, within the treatment of the absorbing boundaries. The practical realisation of this is more difficult than the present case because the required inverse Laplace transform is not of a simple form. The current approach being developed is to use the method in \cite{Jiang2001,CPA:CPA20200,springerlink:10.1007/s10915-012-9620-9} to approximate the more complex inverse Laplace transform.

It should also be possible to reduce the time taken to perform the boundary calculation.  In the oxygen tests it was shown that most of the expense comes from the end of the calculation where the non-locality in time plays a part. One solution to this would be to use the method described in \cite{Jiang2004955} which uses a sum of exponentials approximation that can be evaluated recursively.  The effect is to reduce the sum in \bref{discreteABC} that requires $\mathcal{O}(N)$ operations to one that requires just $\mathcal{O}(\ln N)$.

\subsection{Conclusion}

We have presented a implementation of a spherically symmetric Hartree-Fock system discretised using a Crank-Nicholson scheme. We also presented the derivation and implementation of an absorbing boundary condition approach to handle the outgoing wave condition. It was shown using a Laplace transform method that it is possible to construct a boundary condition at a finite distance away from the origin. This came at the cost of it being non-local in time, meaning the value of the wave-function at the boundary has to be stored throughout the calculation, causing an increase in the time taken to calculate each iteration as it progressed.

The results of the testing show that absorbing boundary conditions do provide a suitable way of treating the boundary in spatially unbounded time-dependent problems. We see that although there are errors introduced from the discretization of the absorbing boundaries, they are small and stay small throughout the various manipulations required to calculate the strength functions. As well as being accurate they also show a good improvement in the speed of the calculation compared to using a large box.

We applied the method to large amplitude motion, and found acceptable results.  We discussed perspectives for future, and more realistic, calculations.

\newpage
\bibliography{tdhf.bib}{}

\begin{thebibliography}{58}%
\makeatletter
\providecommand \@ifxundefined [1]{%
 \@ifx{#1\undefined}
}%
\providecommand \@ifnum [1]{%
 \ifnum #1\expandafter \@firstoftwo
 \else \expandafter \@secondoftwo
 \fi
}%
\providecommand \@ifx [1]{%
 \ifx #1\expandafter \@firstoftwo
 \else \expandafter \@secondoftwo
 \fi
}%
\providecommand \natexlab [1]{#1}%
\providecommand \enquote  [1]{``#1''}%
\providecommand \bibnamefont  [1]{#1}%
\providecommand \bibfnamefont [1]{#1}%
\providecommand \citenamefont [1]{#1}%
\providecommand \href@noop [0]{\@secondoftwo}%
\providecommand \href [0]{\begingroup \@sanitize@url \@href}%
\providecommand \@href[1]{\@@startlink{#1}\@@href}%
\providecommand \@@href[1]{\endgroup#1\@@endlink}%
\providecommand \@sanitize@url [0]{\catcode `\\12\catcode `\$12\catcode
  `\&12\catcode `\#12\catcode `\^12\catcode `\_12\catcode `\%12\relax}%
\providecommand \@@startlink[1]{}%
\providecommand \@@endlink[0]{}%
\providecommand \url  [0]{\begingroup\@sanitize@url \@url }%
\providecommand \@url [1]{\endgroup\@href {#1}{\urlprefix }}%
\providecommand \urlprefix  [0]{URL }%
\providecommand \Eprint [0]{\href }%
\providecommand \doibase [0]{http://dx.doi.org/}%
\providecommand \selectlanguage [0]{\@gobble}%
\providecommand \bibinfo  [0]{\@secondoftwo}%
\providecommand \bibfield  [0]{\@secondoftwo}%
\providecommand \translation [1]{[#1]}%
\providecommand \BibitemOpen [0]{}%
\providecommand \bibitemStop [0]{}%
\providecommand \bibitemNoStop [0]{.\EOS\space}%
\providecommand \EOS [0]{\spacefactor3000\relax}%
\providecommand \BibitemShut  [1]{\csname bibitem#1\endcsname}%
\let\auto@bib@innerbib\@empty
\bibitem [{\citenamefont {Heinen}\ and\ \citenamefont
  {Kull}(2010)}]{springerlink:10.1134/S1054660X10050063}%
  \BibitemOpen
  \bibfield  {author} {\bibinfo {author} {\bibfnamefont {M.}~\bibnamefont
  {Heinen}}\ and\ \bibinfo {author} {\bibfnamefont {H.-J.}\ \bibnamefont
  {Kull}},\ }\href {\doibase 10.1134/S1054660X10050063} {\bibfield  {journal}
  {\bibinfo  {journal} {Laser Physics}\ }\textbf {\bibinfo {volume} {20}},\
  \bibinfo {pages} {581} (\bibinfo {year} {2010})}\BibitemShut {NoStop}%
\bibitem [{\citenamefont {Heinen}\ and\ \citenamefont {Kull}(2009)}]{symABC}%
  \BibitemOpen
  \bibfield  {author} {\bibinfo {author} {\bibfnamefont {M.}~\bibnamefont
  {Heinen}}\ and\ \bibinfo {author} {\bibfnamefont {H.-J.}\ \bibnamefont
  {Kull}},\ }\href {\doibase 10.1103/PhysRevE.79.056709} {\bibfield  {journal}
  {\bibinfo  {journal} {Phys. Rev. E}\ }\textbf {\bibinfo {volume} {79}},\
  \bibinfo {pages} {056709} (\bibinfo {year} {2009})}\BibitemShut {NoStop}%
\bibitem [{\citenamefont {Lau}(2004)}]{Lau2004376}%
  \BibitemOpen
  \bibfield  {author} {\bibinfo {author} {\bibfnamefont {S.~R.}\ \bibnamefont
  {Lau}},\ }\href {\doibase 10.1016/j.jcp.2004.05.013} {\bibfield  {journal}
  {\bibinfo  {journal} {Journal of Computational Physics}\ }\textbf {\bibinfo
  {volume} {199}},\ \bibinfo {pages} {376 } (\bibinfo {year}
  {2004})}\BibitemShut {NoStop}%
\bibitem [{\citenamefont {Alpert}\ \emph {et~al.}(2000)\citenamefont {Alpert},
  \citenamefont {Greengard},\ and\ \citenamefont {Hagstrom}}]{alpert2000}%
  \BibitemOpen
  \bibfield  {author} {\bibinfo {author} {\bibfnamefont {B.}~\bibnamefont
  {Alpert}}, \bibinfo {author} {\bibfnamefont {L.}~\bibnamefont {Greengard}}, \
  and\ \bibinfo {author} {\bibfnamefont {T.}~\bibnamefont {Hagstrom}},\ }\href
  {http://www.jstor.org/stable/2587298} {\bibfield  {journal} {\bibinfo
  {journal} {SIAM Journal on Numerical Analysis}\ }\textbf {\bibinfo {volume}
  {37}},\ \bibinfo {pages} {1138} (\bibinfo {year} {2000})}\BibitemShut
  {NoStop}%
\bibitem [{\citenamefont {Harakeh}\ and\ \citenamefont {van~der
  Woude}(2001{\natexlab{a}})}]{giantResFHMNE}%
  \BibitemOpen
  \bibfield  {author} {\bibinfo {author} {\bibfnamefont {M.~N.}\ \bibnamefont
  {Harakeh}}\ and\ \bibinfo {author} {\bibfnamefont {A.}~\bibnamefont {van~der
  Woude}},\ }\href@noop {} {\emph {\bibinfo {title} {Giant Resonances:
  Fundamental High-Frequency Modes of Nuclear Excitation}}}\ (\bibinfo
  {publisher} {Oxford University Press},\ \bibinfo {address} {Oxford},\
  \bibinfo {year} {2001})\BibitemShut {NoStop}%
\bibitem [{\citenamefont {Reinhard}\ \emph {et~al.}(2006)\citenamefont
  {Reinhard}, \citenamefont {Stevenson}, \citenamefont {Almehed}, \citenamefont
  {Maruhn},\ and\ \citenamefont {Strayer}}]{giantRes}%
  \BibitemOpen
  \bibfield  {author} {\bibinfo {author} {\bibfnamefont {P.-G.}\ \bibnamefont
  {Reinhard}}, \bibinfo {author} {\bibfnamefont {P.~D.}\ \bibnamefont
  {Stevenson}}, \bibinfo {author} {\bibfnamefont {D.}~\bibnamefont {Almehed}},
  \bibinfo {author} {\bibfnamefont {J.~A.}\ \bibnamefont {Maruhn}}, \ and\
  \bibinfo {author} {\bibfnamefont {M.~R.}\ \bibnamefont {Strayer}},\ }\href
  {\doibase 10.1103/PhysRevE.73.036709} {\bibfield  {journal} {\bibinfo
  {journal} {Phys. Rev. E}\ }\textbf {\bibinfo {volume} {73}},\ \bibinfo
  {pages} {036709} (\bibinfo {year} {2006})}\BibitemShut {NoStop}%
\bibitem [{\citenamefont {Nakatsukasa}\ and\ \citenamefont
  {Yabana}(2005{\natexlab{a}})}]{springerlink:10.1140/epjad/i2005-06-052-x}%
  \BibitemOpen
  \bibfield  {author} {\bibinfo {author} {\bibfnamefont {T.}~\bibnamefont
  {Nakatsukasa}}\ and\ \bibinfo {author} {\bibfnamefont {K.}~\bibnamefont
  {Yabana}},\ }\href {\doibase 10.1140/epjad/i2005-06-052-x} {\bibfield
  {journal} {\bibinfo  {journal} {The European Physical Journal A}\ }\textbf
  {\bibinfo {volume} {25}},\ \bibinfo {pages} {527} (\bibinfo {year}
  {2005}{\natexlab{a}})}\BibitemShut {NoStop}%
\bibitem [{\citenamefont {Wu}\ \emph {et~al.}(1999{\natexlab{a}})\citenamefont
  {Wu}, \citenamefont {Strayer},\ and\ \citenamefont
  {Baranger}}]{PhysRevC.60.044302}%
  \BibitemOpen
  \bibfield  {author} {\bibinfo {author} {\bibfnamefont {J.-S.}\ \bibnamefont
  {Wu}}, \bibinfo {author} {\bibfnamefont {M.~R.}\ \bibnamefont {Strayer}}, \
  and\ \bibinfo {author} {\bibfnamefont {M.}~\bibnamefont {Baranger}},\ }\href
  {\doibase 10.1103/PhysRevC.60.044302} {\bibfield  {journal} {\bibinfo
  {journal} {Phys. Rev. C}\ }\textbf {\bibinfo {volume} {60}},\ \bibinfo
  {pages} {044302} (\bibinfo {year} {1999}{\natexlab{a}})}\BibitemShut
  {NoStop}%
\bibitem [{\citenamefont {Antoine}\ and\ \citenamefont
  {Besse}(2003)}]{Antoine2003157}%
  \BibitemOpen
  \bibfield  {author} {\bibinfo {author} {\bibfnamefont {X.}~\bibnamefont
  {Antoine}}\ and\ \bibinfo {author} {\bibfnamefont {C.}~\bibnamefont
  {Besse}},\ }\href {\doibase 10.1016/S0021-9991(03)00159-1} {\bibfield
  {journal} {\bibinfo  {journal} {Journal of Computational Physics}\ }\textbf
  {\bibinfo {volume} {188}},\ \bibinfo {pages} {157 } (\bibinfo {year}
  {2003})}\BibitemShut {NoStop}%
\bibitem [{\citenamefont {Mangin-Brinet}\ \emph {et~al.}(1998)\citenamefont
  {Mangin-Brinet}, \citenamefont {Carbonell},\ and\ \citenamefont
  {Gignoux}}]{1dABC}%
  \BibitemOpen
  \bibfield  {author} {\bibinfo {author} {\bibfnamefont {M.}~\bibnamefont
  {Mangin-Brinet}}, \bibinfo {author} {\bibfnamefont {J.}~\bibnamefont
  {Carbonell}}, \ and\ \bibinfo {author} {\bibfnamefont {C.}~\bibnamefont
  {Gignoux}},\ }\href {\doibase 10.1103/PhysRevA.57.3245} {\bibfield  {journal}
  {\bibinfo  {journal} {Phys. Rev. A}\ }\textbf {\bibinfo {volume} {57}},\
  \bibinfo {pages} {3245} (\bibinfo {year} {1998})}\BibitemShut {NoStop}%
\bibitem [{\citenamefont {Antoine}\ \emph {et~al.}(2008)\citenamefont
  {Antoine}, \citenamefont {Arnold}, \citenamefont {Besse}, \citenamefont
  {Ehrhardt},\ and\ \citenamefont {Schädle}}]{abcreview}%
  \BibitemOpen
  \bibfield  {author} {\bibinfo {author} {\bibfnamefont {X.}~\bibnamefont
  {Antoine}}, \bibinfo {author} {\bibfnamefont {A.}~\bibnamefont {Arnold}},
  \bibinfo {author} {\bibfnamefont {C.}~\bibnamefont {Besse}}, \bibinfo
  {author} {\bibfnamefont {M.}~\bibnamefont {Ehrhardt}}, \ and\ \bibinfo
  {author} {\bibfnamefont {A.}~\bibnamefont {Schädle}},\ }\href@noop {}
  {\bibfield  {journal} {\bibinfo  {journal} {Commun. Comput. Phys.}\ }\textbf
  {\bibinfo {volume} {4}},\ \bibinfo {pages} {729} (\bibinfo {year}
  {2008})}\BibitemShut {NoStop}%
\bibitem [{\citenamefont {Bertsch}\ and\ \citenamefont
  {Broglia}(1994)}]{bertschbroglia}%
  \BibitemOpen
  \bibfield  {author} {\bibinfo {author} {\bibfnamefont {G.~F.}\ \bibnamefont
  {Bertsch}}\ and\ \bibinfo {author} {\bibfnamefont {R.~A.}\ \bibnamefont
  {Broglia}},\ }\href@noop {} {\emph {\bibinfo {title} {Oscillations in Finite
  Quantum Systems}}}\ (\bibinfo  {publisher} {Cambridge University Press},\
  \bibinfo {address} {Cambridge},\ \bibinfo {year} {1994})\BibitemShut
  {NoStop}%
\bibitem [{\citenamefont {Harakeh}\ and\ \citenamefont {van~der
  Woude}(2001{\natexlab{b}})}]{harakeh}%
  \BibitemOpen
  \bibfield  {author} {\bibinfo {author} {\bibfnamefont {M.~N.}\ \bibnamefont
  {Harakeh}}\ and\ \bibinfo {author} {\bibfnamefont {A.}~\bibnamefont {van~der
  Woude}},\ }\href@noop {} {\emph {\bibinfo {title} {Giant Resonances}}}\
  (\bibinfo  {publisher} {Oxford University Press},\ \bibinfo {address}
  {Oxford},\ \bibinfo {year} {2001})\BibitemShut {NoStop}%
\bibitem [{\citenamefont {Harakeh}\ \emph {et~al.}(1977)\citenamefont
  {Harakeh}, \citenamefont {van~der Borg}, \citenamefont {Ishimatsu},
  \citenamefont {Morsch}, \citenamefont {van~der Woude},\ and\ \citenamefont
  {Bertrand}}]{Harakeh1977}%
  \BibitemOpen
  \bibfield  {author} {\bibinfo {author} {\bibfnamefont {M.}~\bibnamefont
  {Harakeh}}, \bibinfo {author} {\bibfnamefont {K.}~\bibnamefont {van~der
  Borg}}, \bibinfo {author} {\bibfnamefont {T.}~\bibnamefont {Ishimatsu}},
  \bibinfo {author} {\bibfnamefont {H.}~\bibnamefont {Morsch}}, \bibinfo
  {author} {\bibfnamefont {A.}~\bibnamefont {van~der Woude}}, \ and\ \bibinfo
  {author} {\bibfnamefont {F.}~\bibnamefont {Bertrand}},\ }\href {\doibase
  10.1103/PhysRevLett.38.676} {\bibfield  {journal} {\bibinfo  {journal}
  {Physical Review Letters}\ }\textbf {\bibinfo {volume} {38}},\ \bibinfo
  {pages} {676} (\bibinfo {year} {1977})}\BibitemShut {NoStop}%
\bibitem [{\citenamefont {Blaizot}(1980)}]{Blaizot1980}%
  \BibitemOpen
  \bibfield  {author} {\bibinfo {author} {\bibfnamefont {J.~P.}\ \bibnamefont
  {Blaizot}},\ }\href@noop {} {\bibfield  {journal} {\bibinfo  {journal}
  {Physics Reports}\ }\textbf {\bibinfo {volume} {4}},\ \bibinfo {pages} {171}
  (\bibinfo {year} {1980})}\BibitemShut {NoStop}%
\bibitem [{\citenamefont {Dutra}\ \emph {et~al.}(2012)\citenamefont {Dutra},
  \citenamefont {Louren\c{c}o}, \citenamefont {{S\'{a} Martins}}, \citenamefont
  {Delfino}, \citenamefont {Stone},\ and\ \citenamefont
  {Stevenson}}]{Dutra2012}%
  \BibitemOpen
  \bibfield  {author} {\bibinfo {author} {\bibfnamefont {M.}~\bibnamefont
  {Dutra}}, \bibinfo {author} {\bibfnamefont {O.}~\bibnamefont {Louren\c{c}o}},
  \bibinfo {author} {\bibfnamefont {J.}~\bibnamefont {{S\'{a} Martins}}},
  \bibinfo {author} {\bibfnamefont {A.}~\bibnamefont {Delfino}}, \bibinfo
  {author} {\bibfnamefont {J.}~\bibnamefont {Stone}}, \ and\ \bibinfo {author}
  {\bibfnamefont {P.~D.}\ \bibnamefont {Stevenson}},\ }\href
  {http://prc.aps.org/abstract/PRC/v85/i3/e035201} {\bibfield  {journal}
  {\bibinfo  {journal} {Physical Review C}\ }\textbf {\bibinfo {volume} {85}},\
  \bibinfo {pages} {035201} (\bibinfo {year} {2012})}\BibitemShut {NoStop}%
\bibitem [{\citenamefont {Vautherin}\ and\ \citenamefont
  {Stringari}(1979)}]{Vautherin1979}%
  \BibitemOpen
  \bibfield  {author} {\bibinfo {author} {\bibfnamefont {D.}~\bibnamefont
  {Vautherin}}\ and\ \bibinfo {author} {\bibfnamefont {S.}~\bibnamefont
  {Stringari}},\ }in\ \href@noop {} {\emph {\bibinfo {booktitle} {Proceedings
  of Workshop on Time-Dependent Hartree-Fock Method, Orsay Saclay May 1979}}},\
  \bibinfo {editor} {edited by\ \bibinfo {editor} {\bibfnamefont
  {P.}~\bibnamefont {Bonche}}}\ (\bibinfo  {publisher} {\'{E}ditions de
  Physique},\ \bibinfo {year} {1979})\ pp.\ \bibinfo {pages}
  {45--49}\BibitemShut {NoStop}%
\bibitem [{\citenamefont {Stringari}\ and\ \citenamefont
  {Vautherin}(1979)}]{Stringari1979}%
  \BibitemOpen
  \bibfield  {author} {\bibinfo {author} {\bibfnamefont {S.}~\bibnamefont
  {Stringari}}\ and\ \bibinfo {author} {\bibfnamefont {D.}~\bibnamefont
  {Vautherin}},\ }\href {\doibase 10.1016/0370-2693(79)90099-6} {\bibfield
  {journal} {\bibinfo  {journal} {Physics Letters B}\ }\textbf {\bibinfo
  {volume} {88}},\ \bibinfo {pages} {1} (\bibinfo {year} {1979})}\BibitemShut
  {NoStop}%
\bibitem [{\citenamefont {Lacroix}\ and\ \citenamefont
  {Chomaz}(1998)}]{Lacroix1998}%
  \BibitemOpen
  \bibfield  {author} {\bibinfo {author} {\bibfnamefont {D.}~\bibnamefont
  {Lacroix}}\ and\ \bibinfo {author} {\bibfnamefont {P.}~\bibnamefont
  {Chomaz}},\ }\href {\doibase 10.1016/S0375-9474(98)00477-1} {\bibfield
  {journal} {\bibinfo  {journal} {Nuclear Physics A}\ }\textbf {\bibinfo
  {volume} {641}},\ \bibinfo {pages} {107} (\bibinfo {year}
  {1998})}\BibitemShut {NoStop}%
\bibitem [{\citenamefont {Wu}\ \emph {et~al.}(1999{\natexlab{b}})\citenamefont
  {Wu}, \citenamefont {Strayer},\ and\ \citenamefont {Baranger}}]{Wu1999}%
  \BibitemOpen
  \bibfield  {author} {\bibinfo {author} {\bibfnamefont {J.-S.}\ \bibnamefont
  {Wu}}, \bibinfo {author} {\bibfnamefont {M.~R.}\ \bibnamefont {Strayer}}, \
  and\ \bibinfo {author} {\bibfnamefont {M.}~\bibnamefont {Baranger}},\ }\href
  {\doibase 10.1103/PhysRevC.60.044302} {\bibfield  {journal} {\bibinfo
  {journal} {Physical Review C}\ }\textbf {\bibinfo {volume} {60}},\ \bibinfo
  {pages} {044302} (\bibinfo {year} {1999}{\natexlab{b}})}\BibitemShut
  {NoStop}%
\bibitem [{\citenamefont {Almehed}\ and\ \citenamefont
  {Stevenson}(2005{\natexlab{a}})}]{Almehed2005}%
  \BibitemOpen
  \bibfield  {author} {\bibinfo {author} {\bibfnamefont {D.}~\bibnamefont
  {Almehed}}\ and\ \bibinfo {author} {\bibfnamefont {P.~D.}\ \bibnamefont
  {Stevenson}},\ }\href {\doibase 10.1063/1.2140673} {\bibfield  {journal}
  {\bibinfo  {journal} {AIP Conference Proceedings}\ }\textbf {\bibinfo
  {volume} {802}},\ \bibinfo {pages} {305} (\bibinfo {year}
  {2005}{\natexlab{a}})}\BibitemShut {NoStop}%
\bibitem [{\citenamefont {Almehed}\ and\ \citenamefont
  {Stevenson}(2005{\natexlab{b}})}]{Almehed2005a}%
  \BibitemOpen
  \bibfield  {author} {\bibinfo {author} {\bibfnamefont {D.}~\bibnamefont
  {Almehed}}\ and\ \bibinfo {author} {\bibfnamefont {P.~D.}\ \bibnamefont
  {Stevenson}},\ }in\ \href {\doibase 10.1088/0954-3899/31/10/079} {\emph
  {\bibinfo {booktitle} {Journal of Physics G}}},\ Vol.~\bibinfo {volume} {31}\
  (\bibinfo {year} {2005})\ pp.\ \bibinfo {pages} {S1819--S1822}\BibitemShut
  {NoStop}%
\bibitem [{\citenamefont {Stevenson}\ and\ \citenamefont
  {Fracasso}(2010)}]{Stevenson2010}%
  \BibitemOpen
  \bibfield  {author} {\bibinfo {author} {\bibfnamefont {P.~D.}\ \bibnamefont
  {Stevenson}}\ and\ \bibinfo {author} {\bibfnamefont {S.}~\bibnamefont
  {Fracasso}},\ }\href {\doibase 10.1088/0954-3899/37/6/064030} {\bibfield
  {journal} {\bibinfo  {journal} {Journal of Physics G: Nuclear and Particle
  Physics}\ }\textbf {\bibinfo {volume} {37}},\ \bibinfo {pages} {064030}
  (\bibinfo {year} {2010})}\BibitemShut {NoStop}%
\bibitem [{\citenamefont {Brandenburg}\ \emph {et~al.}(1983)\citenamefont
  {Brandenburg}, \citenamefont {{De Leo}}, \citenamefont {Drentje},
  \citenamefont {Harakeh}, \citenamefont {Sakai},\ and\ \citenamefont {van~der
  Woude}}]{Brandenburg1983}%
  \BibitemOpen
  \bibfield  {author} {\bibinfo {author} {\bibfnamefont {S.}~\bibnamefont
  {Brandenburg}}, \bibinfo {author} {\bibfnamefont {R.}~\bibnamefont {{De
  Leo}}}, \bibinfo {author} {\bibfnamefont {A.}~\bibnamefont {Drentje}},
  \bibinfo {author} {\bibfnamefont {M.}~\bibnamefont {Harakeh}}, \bibinfo
  {author} {\bibfnamefont {H.}~\bibnamefont {Sakai}}, \ and\ \bibinfo {author}
  {\bibfnamefont {A.}~\bibnamefont {van~der Woude}},\ }\href {\doibase
  10.1016/0370-2693(83)91052-3} {\bibfield  {journal} {\bibinfo  {journal}
  {Physics Letters B}\ }\textbf {\bibinfo {volume} {130}},\ \bibinfo {pages}
  {9} (\bibinfo {year} {1983})}\BibitemShut {NoStop}%
\bibitem [{\citenamefont {Tornow}\ \emph {et~al.}(2012)\citenamefont {Tornow},
  \citenamefont {Kelley}, \citenamefont {Raut}, \citenamefont {Rusev},
  \citenamefont {Tonchev}, \citenamefont {Ahmed}, \citenamefont {Crowell},\
  and\ \citenamefont {Stave}}]{Tornow2012}%
  \BibitemOpen
  \bibfield  {author} {\bibinfo {author} {\bibfnamefont {W.}~\bibnamefont
  {Tornow}}, \bibinfo {author} {\bibfnamefont {J.}~\bibnamefont {Kelley}},
  \bibinfo {author} {\bibfnamefont {R.}~\bibnamefont {Raut}}, \bibinfo {author}
  {\bibfnamefont {G.}~\bibnamefont {Rusev}}, \bibinfo {author} {\bibfnamefont
  {A.}~\bibnamefont {Tonchev}}, \bibinfo {author} {\bibfnamefont
  {M.}~\bibnamefont {Ahmed}}, \bibinfo {author} {\bibfnamefont
  {A.}~\bibnamefont {Crowell}}, \ and\ \bibinfo {author} {\bibfnamefont
  {S.}~\bibnamefont {Stave}},\ }\href
  {http://prc.aps.org/abstract/PRC/v85/i6/e061001} {\bibfield  {journal}
  {\bibinfo  {journal} {Physical Review C}\ }\textbf {\bibinfo {volume} {85}},\
  \bibinfo {pages} {061001(R)} (\bibinfo {year} {2012})}\BibitemShut {NoStop}%
\bibitem [{\citenamefont {Pines}\ and\ \citenamefont
  {Nozi\`eres}(1966)}]{NozPines}%
  \BibitemOpen
  \bibfield  {author} {\bibinfo {author} {\bibfnamefont {D.}~\bibnamefont
  {Pines}}\ and\ \bibinfo {author} {\bibfnamefont {P.}~\bibnamefont
  {Nozi\`eres}},\ }\href@noop {} {\emph {\bibinfo {title} {The Theory of
  Quantum Liquids, 1: Normal Fermi Liquids}}},\ \bibinfo {edition} {1st}\ ed.\
  (\bibinfo  {publisher} {W. A. Benjamin, Inc.},\ \bibinfo {address} {New
  York},\ \bibinfo {year} {1966})\BibitemShut {NoStop}%
\bibitem [{\citenamefont {Chinn}\ \emph {et~al.}(1996)\citenamefont {Chinn},
  \citenamefont {Umar}, \citenamefont {Valli\`{e}res},\ and\ \citenamefont
  {Strayer}}]{Chinn1996}%
  \BibitemOpen
  \bibfield  {author} {\bibinfo {author} {\bibfnamefont {C.}~\bibnamefont
  {Chinn}}, \bibinfo {author} {\bibfnamefont {A.}~\bibnamefont {Umar}},
  \bibinfo {author} {\bibfnamefont {M.}~\bibnamefont {Valli\`{e}res}}, \ and\
  \bibinfo {author} {\bibfnamefont {M.}~\bibnamefont {Strayer}},\ }\href
  {http://dx.doi.org/10.1016/0370-1573(95)00031-3} {\bibfield  {journal}
  {\bibinfo  {journal} {Physics Reports}\ }\textbf {\bibinfo {volume} {264}},\
  \bibinfo {pages} {107} (\bibinfo {year} {1996})}\BibitemShut {NoStop}%
\bibitem [{\citenamefont {Dirac}(1930)}]{Dirac1930}%
  \BibitemOpen
  \bibfield  {author} {\bibinfo {author} {\bibfnamefont {P.~A.~M.}\
  \bibnamefont {Dirac}},\ }\href@noop {} {\bibfield  {journal} {\bibinfo
  {journal} {Proceedings of the Cambridge Philosophical Society}\ }\textbf
  {\bibinfo {volume} {24}},\ \bibinfo {pages} {376} (\bibinfo {year}
  {1930})}\BibitemShut {NoStop}%
\bibitem [{\citenamefont {Bonche}\ \emph {et~al.}(1976)\citenamefont {Bonche},
  \citenamefont {Koonin},\ and\ \citenamefont {Negele}}]{Bonche1976}%
  \BibitemOpen
  \bibfield  {author} {\bibinfo {author} {\bibfnamefont {P.}~\bibnamefont
  {Bonche}}, \bibinfo {author} {\bibfnamefont {S.}~\bibnamefont {Koonin}}, \
  and\ \bibinfo {author} {\bibfnamefont {J.}~\bibnamefont {Negele}},\ }\href
  {\doibase 10.1103/PhysRevC.13.1226} {\bibfield  {journal} {\bibinfo
  {journal} {Physical Review C}\ }\textbf {\bibinfo {volume} {13}},\ \bibinfo
  {pages} {1226} (\bibinfo {year} {1976})}\BibitemShut {NoStop}%
\bibitem [{\citenamefont {Cusson}\ \emph {et~al.}(1976)\citenamefont {Cusson},
  \citenamefont {Smith},\ and\ \citenamefont {Maruhn}}]{Cusson1976}%
  \BibitemOpen
  \bibfield  {author} {\bibinfo {author} {\bibfnamefont {R.}~\bibnamefont
  {Cusson}}, \bibinfo {author} {\bibfnamefont {R.}~\bibnamefont {Smith}}, \
  and\ \bibinfo {author} {\bibfnamefont {J.}~\bibnamefont {Maruhn}},\ }\href
  {\doibase 10.1103/PhysRevLett.36.1166} {\bibfield  {journal} {\bibinfo
  {journal} {Physical Review Letters}\ }\textbf {\bibinfo {volume} {36}},\
  \bibinfo {pages} {1166} (\bibinfo {year} {1976})}\BibitemShut {NoStop}%
\bibitem [{\citenamefont {Devi}\ \emph {et~al.}(1979)\citenamefont {Devi},
  \citenamefont {Strayer},\ and\ \citenamefont {Irvine}}]{Devi1979}%
  \BibitemOpen
  \bibfield  {author} {\bibinfo {author} {\bibfnamefont {K.~R.~S.}\
  \bibnamefont {Devi}}, \bibinfo {author} {\bibfnamefont {M.~R.}\ \bibnamefont
  {Strayer}}, \ and\ \bibinfo {author} {\bibfnamefont {J.~M.}\ \bibnamefont
  {Irvine}},\ }\href {\doibase 10.1088/0305-4616/5/2/015} {\bibfield  {journal}
  {\bibinfo  {journal} {Journal of Physics G: Nuclear Physics}\ }\textbf
  {\bibinfo {volume} {5}},\ \bibinfo {pages} {281} (\bibinfo {year}
  {1979})}\BibitemShut {NoStop}%
\bibitem [{\citenamefont {Maruhn}\ \emph {et~al.}(2006)\citenamefont {Maruhn},
  \citenamefont {Reinhard}, \citenamefont {Stevenson},\ and\ \citenamefont
  {Strayer}}]{Maruhn2006}%
  \BibitemOpen
  \bibfield  {author} {\bibinfo {author} {\bibfnamefont {J.~A.}\ \bibnamefont
  {Maruhn}}, \bibinfo {author} {\bibfnamefont {P.-G.}\ \bibnamefont
  {Reinhard}}, \bibinfo {author} {\bibfnamefont {P.~D.}\ \bibnamefont
  {Stevenson}}, \ and\ \bibinfo {author} {\bibfnamefont {M.~R.}\ \bibnamefont
  {Strayer}},\ }\href {\doibase 10.1103/PhysRevC.74.027601} {\bibfield
  {journal} {\bibinfo  {journal} {Physical Review C}\ }\textbf {\bibinfo
  {volume} {74}},\ \bibinfo {pages} {027601} (\bibinfo {year}
  {2006})}\BibitemShut {NoStop}%
\bibitem [{\citenamefont {Stevenson}\ \emph {et~al.}(2004)\citenamefont
  {Stevenson}, \citenamefont {Strayer}, \citenamefont {Stone},\ and\
  \citenamefont {Newton}}]{Stevenson2004}%
  \BibitemOpen
  \bibfield  {author} {\bibinfo {author} {\bibfnamefont {P.~D.}\ \bibnamefont
  {Stevenson}}, \bibinfo {author} {\bibfnamefont {M.~R.}\ \bibnamefont
  {Strayer}}, \bibinfo {author} {\bibfnamefont {J.~R.}\ \bibnamefont {Stone}},
  \ and\ \bibinfo {author} {\bibfnamefont {W.~G.}\ \bibnamefont {Newton}},\
  }\href@noop {} {\bibfield  {journal} {\bibinfo  {journal} {International
  Journal of Modern Physics E}\ }\textbf {\bibinfo {volume} {13}},\ \bibinfo
  {pages} {181} (\bibinfo {year} {2004})}\BibitemShut {NoStop}%
\bibitem [{\citenamefont {Umar}\ and\ \citenamefont
  {Oberacker}(2006)}]{Umar2006}%
  \BibitemOpen
  \bibfield  {author} {\bibinfo {author} {\bibfnamefont {A.~S.}\ \bibnamefont
  {Umar}}\ and\ \bibinfo {author} {\bibfnamefont {V.}~\bibnamefont
  {Oberacker}},\ }\href {\doibase 10.1103/PhysRevC.73.054607} {\bibfield
  {journal} {\bibinfo  {journal} {Physical Review C}\ }\textbf {\bibinfo
  {volume} {73}},\ \bibinfo {pages} {7} (\bibinfo {year} {2006})}\BibitemShut
  {NoStop}%
\bibitem [{\citenamefont {Fracasso}\ \emph {et~al.}(2012)\citenamefont
  {Fracasso}, \citenamefont {Suckling},\ and\ \citenamefont
  {Stevenson}}]{PhysRevC.86.044303}%
  \BibitemOpen
  \bibfield  {author} {\bibinfo {author} {\bibfnamefont {S.}~\bibnamefont
  {Fracasso}}, \bibinfo {author} {\bibfnamefont {E.~B.}\ \bibnamefont
  {Suckling}}, \ and\ \bibinfo {author} {\bibfnamefont {P.~D.}\ \bibnamefont
  {Stevenson}},\ }\href {\doibase 10.1103/PhysRevC.86.044303} {\bibfield
  {journal} {\bibinfo  {journal} {Phys. Rev. C}\ }\textbf {\bibinfo {volume}
  {86}},\ \bibinfo {pages} {044303} (\bibinfo {year} {2012})}\BibitemShut
  {NoStop}%
\bibitem [{\citenamefont {Engel}\ \emph {et~al.}(1975)\citenamefont {Engel},
  \citenamefont {Brink}, \citenamefont {Goeke}, \citenamefont {Krieger},\ and\
  \citenamefont {Vautherin}}]{Engel1975}%
  \BibitemOpen
  \bibfield  {author} {\bibinfo {author} {\bibfnamefont {Y.~M.}\ \bibnamefont
  {Engel}}, \bibinfo {author} {\bibfnamefont {D.~M.}\ \bibnamefont {Brink}},
  \bibinfo {author} {\bibfnamefont {K.}~\bibnamefont {Goeke}}, \bibinfo
  {author} {\bibfnamefont {J.}~\bibnamefont {Krieger}}, \ and\ \bibinfo
  {author} {\bibfnamefont {D.}~\bibnamefont {Vautherin}},\ }\href@noop {}
  {\bibfield  {journal} {\bibinfo  {journal} {Nuclear Physics}\ }\textbf
  {\bibinfo {volume} {249}},\ \bibinfo {pages} {215} (\bibinfo {year}
  {1975})}\BibitemShut {NoStop}%
\bibitem [{\citenamefont {Wu}\ \emph {et~al.}(1997)\citenamefont {Wu},
  \citenamefont {Wong}, \citenamefont {Strayer},\ and\ \citenamefont
  {Baranger}}]{PhysRevC.56.857}%
  \BibitemOpen
  \bibfield  {author} {\bibinfo {author} {\bibfnamefont {J.-S.}\ \bibnamefont
  {Wu}}, \bibinfo {author} {\bibfnamefont {K.~C.}\ \bibnamefont {Wong}},
  \bibinfo {author} {\bibfnamefont {M.~R.}\ \bibnamefont {Strayer}}, \ and\
  \bibinfo {author} {\bibfnamefont {M.}~\bibnamefont {Baranger}},\ }\href
  {\doibase 10.1103/PhysRevC.56.857} {\bibfield  {journal} {\bibinfo  {journal}
  {Phys. Rev. C}\ }\textbf {\bibinfo {volume} {56}},\ \bibinfo {pages} {857}
  (\bibinfo {year} {1997})}\BibitemShut {NoStop}%
\bibitem [{\citenamefont {Nakatsukasa}\ and\ \citenamefont
  {Yabana}(2005{\natexlab{b}})}]{PhysRevC.71.024301}%
  \BibitemOpen
  \bibfield  {author} {\bibinfo {author} {\bibfnamefont {T.}~\bibnamefont
  {Nakatsukasa}}\ and\ \bibinfo {author} {\bibfnamefont {K.}~\bibnamefont
  {Yabana}},\ }\href {\doibase 10.1103/PhysRevC.71.024301} {\bibfield
  {journal} {\bibinfo  {journal} {Phys. Rev. C}\ }\textbf {\bibinfo {volume}
  {71}},\ \bibinfo {pages} {024301} (\bibinfo {year}
  {2005}{\natexlab{b}})}\BibitemShut {NoStop}%
\bibitem [{\citenamefont {Abramowitz}\ and\ \citenamefont
  {Stegun}(1965)}]{AbraMathFunc}%
  \BibitemOpen
  \bibfield  {author} {\bibinfo {author} {\bibfnamefont {M.}~\bibnamefont
  {Abramowitz}}\ and\ \bibinfo {author} {\bibfnamefont {I.~A.}\ \bibnamefont
  {Stegun}},\ }\href@noop {} {\emph {\bibinfo {title} {Handbook of Mathematical
  Functions: with Formulas, Graphs, and Mathematical Tables}}},\ \bibinfo
  {edition} {3rd}\ ed.\ (\bibinfo  {publisher} {Academic Press},\ \bibinfo
  {year} {1965})\BibitemShut {NoStop}%
\bibitem [{\citenamefont {Duffy}(2004)}]{transMethDuffy}%
  \BibitemOpen
  \bibfield  {author} {\bibinfo {author} {\bibfnamefont {D.~G.}\ \bibnamefont
  {Duffy}},\ }\href@noop {} {\emph {\bibinfo {title} {Transform Methods for
  Solving Partial Differential Equations}}},\ \bibinfo {edition} {2nd}\ ed.\
  (\bibinfo  {publisher} {Chapman and Hall},\ \bibinfo {year}
  {2004})\BibitemShut {NoStop}%
\bibitem [{\citenamefont {Erdelyi}(1954)}]{intTransErdelyi}%
  \BibitemOpen
  \bibfield  {author} {\bibinfo {author} {\bibfnamefont {A.}~\bibnamefont
  {Erdelyi}},\ }\href@noop {} {\emph {\bibinfo {title} {Tables of Integral
  Transforms}}}\ (\bibinfo  {publisher} {McGraw Hill Text},\ \bibinfo {year}
  {1954})\BibitemShut {NoStop}%
\bibitem [{\citenamefont {Poppe}\ and\ \citenamefont
  {Wijers}(1990)}]{Poppe:1990:MEC:77626.77629}%
  \BibitemOpen
  \bibfield  {author} {\bibinfo {author} {\bibfnamefont {G.~P.~M.}\
  \bibnamefont {Poppe}}\ and\ \bibinfo {author} {\bibfnamefont {C.~M.~J.}\
  \bibnamefont {Wijers}},\ }\href {\doibase 10.1145/77626.77629} {\bibfield
  {journal} {\bibinfo  {journal} {ACM Trans. Math. Softw.}\ }\textbf {\bibinfo
  {volume} {16}},\ \bibinfo {pages} {38} (\bibinfo {year} {1990})}\BibitemShut
  {NoStop}%
\bibitem [{\citenamefont {Boucke}\ \emph {et~al.}(1997)\citenamefont {Boucke},
  \citenamefont {Schmitz},\ and\ \citenamefont {Kull}}]{symABCprev}%
  \BibitemOpen
  \bibfield  {author} {\bibinfo {author} {\bibfnamefont {K.}~\bibnamefont
  {Boucke}}, \bibinfo {author} {\bibfnamefont {H.}~\bibnamefont {Schmitz}}, \
  and\ \bibinfo {author} {\bibfnamefont {H.-J.}\ \bibnamefont {Kull}},\ }\href
  {\doibase 10.1103/PhysRevA.56.763} {\bibfield  {journal} {\bibinfo  {journal}
  {Phys. Rev. A}\ }\textbf {\bibinfo {volume} {56}},\ \bibinfo {pages} {763}
  (\bibinfo {year} {1997})}\BibitemShut {NoStop}%
\bibitem [{\citenamefont {Stevenson}\ \emph {et~al.}(2007)\citenamefont
  {Stevenson}, \citenamefont {Almehed}, \citenamefont {Reinhard},\ and\
  \citenamefont {Maruhn}}]{Stevenson2007}%
  \BibitemOpen
  \bibfield  {author} {\bibinfo {author} {\bibfnamefont {P.~D.}\ \bibnamefont
  {Stevenson}}, \bibinfo {author} {\bibfnamefont {D.}~\bibnamefont {Almehed}},
  \bibinfo {author} {\bibfnamefont {P.-G.}\ \bibnamefont {Reinhard}}, \ and\
  \bibinfo {author} {\bibfnamefont {J.~A.}\ \bibnamefont {Maruhn}},\ }\href
  {\doibase 10.1016/j.nuclphysa.2007.01.091} {\bibfield  {journal} {\bibinfo
  {journal} {Nuclear Physics A}\ }\textbf {\bibinfo {volume} {788}},\ \bibinfo
  {pages} {343} (\bibinfo {year} {2007})}\BibitemShut {NoStop}%
\bibitem [{\citenamefont {Simenel}\ and\ \citenamefont
  {Chomaz}(2003)}]{PhysRevC.68.024302}%
  \BibitemOpen
  \bibfield  {author} {\bibinfo {author} {\bibfnamefont {C.}~\bibnamefont
  {Simenel}}\ and\ \bibinfo {author} {\bibfnamefont {P.}~\bibnamefont
  {Chomaz}},\ }\href {\doibase 10.1103/PhysRevC.68.024302} {\bibfield
  {journal} {\bibinfo  {journal} {Phys. Rev. C}\ }\textbf {\bibinfo {volume}
  {68}},\ \bibinfo {pages} {024302} (\bibinfo {year} {2003})}\BibitemShut
  {NoStop}%
\bibitem [{\citenamefont {Simenel}\ and\ \citenamefont
  {Chomaz}(2009)}]{PhysRevC.80.064309}%
  \BibitemOpen
  \bibfield  {author} {\bibinfo {author} {\bibfnamefont {C.}~\bibnamefont
  {Simenel}}\ and\ \bibinfo {author} {\bibfnamefont {P.}~\bibnamefont
  {Chomaz}},\ }\href {\doibase 10.1103/PhysRevC.80.064309} {\bibfield
  {journal} {\bibinfo  {journal} {Phys. Rev. C}\ }\textbf {\bibinfo {volume}
  {80}},\ \bibinfo {pages} {064309} (\bibinfo {year} {2009})}\BibitemShut
  {NoStop}%
\bibitem [{\citenamefont {Reinhard}\ \emph {et~al.}(2007)\citenamefont
  {Reinhard}, \citenamefont {Guo},\ and\ \citenamefont
  {Maruhn}}]{Reinhard2007}%
  \BibitemOpen
  \bibfield  {author} {\bibinfo {author} {\bibfnamefont {P.~G.}\ \bibnamefont
  {Reinhard}}, \bibinfo {author} {\bibfnamefont {L.}~\bibnamefont {Guo}}, \
  and\ \bibinfo {author} {\bibfnamefont {J.~A.}\ \bibnamefont {Maruhn}},\
  }\href {\doibase 10.1140/epja/i2007-10366-9} {\bibfield  {journal} {\bibinfo
  {journal} {The European Physical Journal A}\ }\textbf {\bibinfo {volume}
  {32}},\ \bibinfo {pages} {19} (\bibinfo {year} {2007})}\BibitemShut {NoStop}%
\bibitem [{\citenamefont {Reinhard}\ and\ \citenamefont
  {Suraud}(2003)}]{cluster}%
  \BibitemOpen
  \bibfield  {author} {\bibinfo {author} {\bibfnamefont {P.-G.}\ \bibnamefont
  {Reinhard}}\ and\ \bibinfo {author} {\bibfnamefont {E.}~\bibnamefont
  {Suraud}},\ }\href@noop {} {\emph {\bibinfo {title} {Introduction to Cluster
  Dynamics}}}\ (\bibinfo  {publisher} {Wiley},\ \bibinfo {address} {Berlin},\
  \bibinfo {year} {2003})\BibitemShut {NoStop}%
\bibitem [{\citenamefont {Maruhn}\ \emph {et~al.}(2005)\citenamefont {Maruhn},
  \citenamefont {Reinhard}, \citenamefont {Stevenson}, \citenamefont {Stone},\
  and\ \citenamefont {Strayer}}]{PhysRevC.71.064328}%
  \BibitemOpen
  \bibfield  {author} {\bibinfo {author} {\bibfnamefont {J.~A.}\ \bibnamefont
  {Maruhn}}, \bibinfo {author} {\bibfnamefont {P.~G.}\ \bibnamefont
  {Reinhard}}, \bibinfo {author} {\bibfnamefont {P.~D.}\ \bibnamefont
  {Stevenson}}, \bibinfo {author} {\bibfnamefont {J.~R.}\ \bibnamefont
  {Stone}}, \ and\ \bibinfo {author} {\bibfnamefont {M.~R.}\ \bibnamefont
  {Strayer}},\ }\href {\doibase 10.1103/PhysRevC.71.064328} {\bibfield
  {journal} {\bibinfo  {journal} {Phys. Rev. C}\ }\textbf {\bibinfo {volume}
  {71}},\ \bibinfo {pages} {064328} (\bibinfo {year} {2005})}\BibitemShut
  {NoStop}%
\bibitem [{\citenamefont {Brine}\ \emph {et~al.}(2006)\citenamefont {Brine},
  \citenamefont {Stevenson}, \citenamefont {Maruhn},\ and\ \citenamefont
  {Reinhard}}]{brinemg}%
  \BibitemOpen
  \bibfield  {author} {\bibinfo {author} {\bibfnamefont {M.~P.}\ \bibnamefont
  {Brine}}, \bibinfo {author} {\bibfnamefont {P.~D.}\ \bibnamefont
  {Stevenson}}, \bibinfo {author} {\bibfnamefont {J.~A.}\ \bibnamefont
  {Maruhn}}, \ and\ \bibinfo {author} {\bibfnamefont {P.-G.}\ \bibnamefont
  {Reinhard}},\ }\href {\doibase 10.1142/S0218301306005009} {\bibfield
  {journal} {\bibinfo  {journal} {International Journal of Modern Physics E}\
  }\textbf {\bibinfo {volume} {15}},\ \bibinfo {pages} {1417} (\bibinfo {year}
  {2006})}\BibitemShut {NoStop}%
\bibitem [{\citenamefont {Umar}\ and\ \citenamefont
  {Oberacker}(2005)}]{Umar2005}%
  \BibitemOpen
  \bibfield  {author} {\bibinfo {author} {\bibfnamefont {A.~S.}\ \bibnamefont
  {Umar}}\ and\ \bibinfo {author} {\bibfnamefont {V.}~\bibnamefont
  {Oberacker}},\ }\href {\doibase 10.1103/PhysRevC.71.034314} {\bibfield
  {journal} {\bibinfo  {journal} {Physical Review C}\ }\textbf {\bibinfo
  {volume} {71}},\ \bibinfo {pages} {034314} (\bibinfo {year}
  {2005})}\BibitemShut {NoStop}%
\bibitem [{\citenamefont {Ebata}\ \emph {et~al.}(2010)\citenamefont {Ebata},
  \citenamefont {Nakatsukasa}, \citenamefont {Inakura}, \citenamefont
  {Yoshida}, \citenamefont {Hashimoto},\ and\ \citenamefont
  {Yabana}}]{PhysRevC.82.034306}%
  \BibitemOpen
  \bibfield  {author} {\bibinfo {author} {\bibfnamefont {S.}~\bibnamefont
  {Ebata}}, \bibinfo {author} {\bibfnamefont {T.}~\bibnamefont {Nakatsukasa}},
  \bibinfo {author} {\bibfnamefont {T.}~\bibnamefont {Inakura}}, \bibinfo
  {author} {\bibfnamefont {K.}~\bibnamefont {Yoshida}}, \bibinfo {author}
  {\bibfnamefont {Y.}~\bibnamefont {Hashimoto}}, \ and\ \bibinfo {author}
  {\bibfnamefont {K.}~\bibnamefont {Yabana}},\ }\href {\doibase
  10.1103/PhysRevC.82.034306} {\bibfield  {journal} {\bibinfo  {journal} {Phys.
  Rev. C}\ }\textbf {\bibinfo {volume} {82}},\ \bibinfo {pages} {034306}
  (\bibinfo {year} {2010})}\BibitemShut {NoStop}%
\bibitem [{\citenamefont {Avez}\ \emph {et~al.}(2008)\citenamefont {Avez},
  \citenamefont {Simenel},\ and\ \citenamefont {Chomaz}}]{PhysRevC.78.044318}%
  \BibitemOpen
  \bibfield  {author} {\bibinfo {author} {\bibfnamefont {B.}~\bibnamefont
  {Avez}}, \bibinfo {author} {\bibfnamefont {C.}~\bibnamefont {Simenel}}, \
  and\ \bibinfo {author} {\bibfnamefont {P.}~\bibnamefont {Chomaz}},\ }\href
  {\doibase 10.1103/PhysRevC.78.044318} {\bibfield  {journal} {\bibinfo
  {journal} {Phys. Rev. C}\ }\textbf {\bibinfo {volume} {78}},\ \bibinfo
  {pages} {044318} (\bibinfo {year} {2008})}\BibitemShut {NoStop}%
\bibitem [{\citenamefont {Stetcu}\ \emph {et~al.}(2011)\citenamefont {Stetcu},
  \citenamefont {Bulgac}, \citenamefont {Magierski},\ and\ \citenamefont
  {Roche}}]{PhysRevC.84.051309}%
  \BibitemOpen
  \bibfield  {author} {\bibinfo {author} {\bibfnamefont {I.}~\bibnamefont
  {Stetcu}}, \bibinfo {author} {\bibfnamefont {A.}~\bibnamefont {Bulgac}},
  \bibinfo {author} {\bibfnamefont {P.}~\bibnamefont {Magierski}}, \ and\
  \bibinfo {author} {\bibfnamefont {K.~J.}\ \bibnamefont {Roche}},\ }\href
  {\doibase 10.1103/PhysRevC.84.051309} {\bibfield  {journal} {\bibinfo
  {journal} {Phys. Rev. C}\ }\textbf {\bibinfo {volume} {84}},\ \bibinfo
  {pages} {051309} (\bibinfo {year} {2011})}\BibitemShut {NoStop}%
\bibitem [{\citenamefont {Jiang}(2001)}]{Jiang2001}%
  \BibitemOpen
  \bibfield  {author} {\bibinfo {author} {\bibfnamefont {S.}~\bibnamefont
  {Jiang}},\ }\emph {\bibinfo {title} {Fast evaluation of the nonreflecting
  boundary conditions for the Schr\"odinger equation.}},\ \href@noop {} {Ph.D.
  thesis},\ \bibinfo  {school} {Courant Institute of Mathematical Sciences, New
  York University.} (\bibinfo {year} {2001})\BibitemShut {NoStop}%
\bibitem [{\citenamefont {Jiang}\ and\ \citenamefont
  {Greengard}(2008)}]{CPA:CPA20200}%
  \BibitemOpen
  \bibfield  {author} {\bibinfo {author} {\bibfnamefont {S.}~\bibnamefont
  {Jiang}}\ and\ \bibinfo {author} {\bibfnamefont {L.}~\bibnamefont
  {Greengard}},\ }\href {\doibase 10.1002/cpa.20200} {\bibfield  {journal}
  {\bibinfo  {journal} {Communications on Pure and Applied Mathematics}\
  }\textbf {\bibinfo {volume} {61}},\ \bibinfo {pages} {261} (\bibinfo {year}
  {2008})}\BibitemShut {NoStop}%
\bibitem [{\citenamefont {Xu}\ and\ \citenamefont
  {Jiang}(2012)}]{springerlink:10.1007/s10915-012-9620-9}%
  \BibitemOpen
  \bibfield  {author} {\bibinfo {author} {\bibfnamefont {K.}~\bibnamefont
  {Xu}}\ and\ \bibinfo {author} {\bibfnamefont {S.}~\bibnamefont {Jiang}},\
  }\href {\doibase 10.1007/s10915-012-9620-9} {\bibfield  {journal} {\bibinfo
  {journal} {Journal of Scientific Computing}\ } (\bibinfo {year} {2012}),\
  10.1007/s10915-012-9620-9}\BibitemShut {NoStop}%
\bibitem [{\citenamefont {Jiang}\ and\ \citenamefont
  {Greengard}(2004)}]{Jiang2004955}%
  \BibitemOpen
  \bibfield  {author} {\bibinfo {author} {\bibfnamefont {S.}~\bibnamefont
  {Jiang}}\ and\ \bibinfo {author} {\bibfnamefont {L.}~\bibnamefont
  {Greengard}},\ }\href {\doibase 10.1016/S0898-1221(04)90079-X} {\bibfield
  {journal} {\bibinfo  {journal} {Computers \& Mathematics with Applications}\
  }\textbf {\bibinfo {volume} {47}},\ \bibinfo {pages} {955 } (\bibinfo {year}
  {2004})}\BibitemShut {NoStop}%
\end{thebibliography}%

\end{document}